\documentclass[11pt,aps,prd,eqsecnum,superscriptaddress,a4paper]{revtex4-2}
\usepackage{amssymb,amsmath,amsthm,graphicx,amscd}
\usepackage{enumerate,comment,ulem,bm,enumitem}
\usepackage[mathscr]{eucal}
\usepackage[cal=boondoxo]{mathalpha}

\usepackage{xcolor}
\usepackage{tikz}
\usetikzlibrary{arrows.meta}

\usepackage[hidelinks]{hyperref}
\usepackage{orcidlink}

\makeatletter
\renewcommand*\env@matrix[1][\arraystretch]{%
  \edef\arraystretch{#1}%
  \hskip -\arraycolsep
  \let\@ifnextchar\new@ifnextchar
  \array{*\c@MaxMatrixCols c}}
 
\makeatother
\begin{document}

\title{Quantum optomechanics with arbitrary mirror displacement:\\ nonlinear Langevin equation with non-Markovian back-action noises}

\author{Adri\'an E. Rubio L\'opez\orcidlink{0000-0002-8369-193X}}
\email{adrianrubiolopez0102@gmail.com}
\affiliation{Department of Physics, Universidad de Santiago de Chile, Av. Victor Jara 3493, Santiago, Chile.}
\affiliation{Millennium Institute for Research in Optics, Concepci\'on, Chile.}

\author{Hing-Tong Cho\orcidlink{0000-0002-8497-1490}}
\email{htcho@mail.tku.edu.tw}
\affiliation{Department of Physics, Tamkang University, Tamsui, Taipei, Taiwan.}

\author{Bei-Lok Hu\orcidlink{0000-0003-2489-9914}}
\email{blhu@umd.edu}
\affiliation{Joint Quantum Institute and Maryland Center for Fundamental Physics, University of Maryland, College Park, Maryland 20742, USA}

\date{Began in mid-April 2026. This version \today}

\begin{abstract}
Quantum optomechanics (QOM) explores the interaction between a quantum field, often confined in a cavity, and the quantum motion of mirrors, membranes or material media, usually assumed slow enough  with negligible particle production, which is the central concern of dynamical Casimir effects, and a drive laser field for effective control.  This rapidly developing field has a wide range of applications, from quantum sensing to the detection of gravitational waves. Because the opto-mechanical coupling is nonlinear, most theoretical investigations assume that the amplitude of mirror motion $x$ remains small. Recent years saw several papers treating $x^2, x^3$ orders in the mirror displacement. In our work we take a different route, deploying the functional perturbative method presented in \cite{HPZ93} and enriched in \cite{ChoHu26}, applicable for sufficiently weak opto-mechanical couplings. With this we can treat arbitrary displacement of the mirror, not just at a higher order in a power series expansion in $x$. We first derive the influence action of the quantum field and its non-Markovian noises  back-reacting on the moving mirror with self-consistency. From it we derive a nonlinear Langevin equation driven by these back-action noises from the quantum field. We then show how results from our general modeling and treatment can be linked with the models and the Langevin equations presented in the literature, starting with the popular radiation pressure $Nx$ coupling, with $N$ the photon number, up to the recent $x^3$ order results. We hope this new approach and the results presented here can provide useful theoretical support for high precision QOM experimentation in the future.  
\end{abstract}


\maketitle

\newpage
\tableofcontents

\section{Introduction}

Cavity-optomechanics \cite{Kip07,AKMrev} involves the interaction between light and the mechanical motion of matter in typical setups consisting of one or several electromagnetic modes confined by a cavity made of two mirrors or walls of highly reflective material, where one of them is assumed to be movable. Depending on the order of displacement and how rapidly the wall moves,  different hamiltonians for the light-matter interaction have been introduced. Quantum optomechanics (QOM) \cite{Mey13,MilburnQOM} treats all parties involved,  mirror motion and cavity field, as dynamical quantum variables, allowing for the proper role of quantum fluctuations both in the mirror displacements and in the quantum field,  which are absent in a classical treatment. Quantum field fluctuations are the root cause of  radiation pressure fluctuations \cite{Caves}, which, in a semiclassical treatment in the popular Fock space representation, involves the photon number and its fluctuations. Their back-action \cite{backaction}, known as backaction noise,  on the mirror's motion is an important factor in the consideration of noise-reduction schemes in ultra-sensitive experiments \cite{CavesRMP}, such as for gravitational wave detection in LIGO/VIRGO/KAGRA \cite{LIGO,VIRGO}. Recent measurements of radiation pressure backaction noise in a macroscopic object have been reported in \cite{Purdy}, and for VIRGO \cite{VirgoNoise} and LIGO \cite{LIGOnoise} experiments.   

The most studied cases in QOM are for a linear displacement $x$ of the mirror moving sufficiently slowly in a single field mode cavity, as a result of the radiation pressure exerted on it proportional to the photon number $N$.  A nonperturbative treatment of this generic case with  $Nx$ type of coupling by numerical means has been reported in Ref.\cite{Macri2018}. Under strong couplings, perturbative treatments are no longer adequate. Rabi oscillations between the cavity's and the mirror's fundamental modes appear and the eigen-space of the whole system need be accounted for, not either of its subsystems. To go beyond the linear displacement $Nx$ case analytically, one natural pathway is to seek a power law expansion in the displacement $x$. Along this route we mention  the work of Bhattacharya et al \cite{Mey08}, Fosco et al \cite{Fosco11}, Khorasani \cite{Kho17,Kho18} and of Butera \cite{ButeraHi} to $x^2$ order, and that of Ferreri et al \cite{Ferreri2025} to the cubic order $x^3$.  

In this work we present an alternative pathway to go  beyond the small displacement restriction. Our methodology can treat the $N f(x)$  type, where $f(x)$ is an arbitrary analytic function of $x$, and more,  because it is based on \cite{ChoHu26} which includes different coupling constants for the couplings with $q^2$ and $p^2$ and the cross terms. The only restriction is weak opto-mechanical interactions. However, as pointed out in \cite{ButeraHi}, a weak coupling approximation need be, and is probably implicitly assumed, when one carries out a power law expansion in the displacement. If otherwise, as mentioned above, strong coupling regime demands a very different treatment and the physics is qualitatively different from weak interactions.  After we have obtained the results for these more general cases  we will then compare our results with other state-of-the-art models (described in more details in the next section) such as Ref. \cite{Ferreri2025} which obtained an expression for the Hamiltonian up to the third order in the wall's displacement. 

In addition to the magnitude of the mirror's displacement, how fast the mirror moves is also a major factor to be reckoned with. Quantum optomechanics works in the regime where particle number (photons for EM field) is an adiabatic invariant, meaning,  the movement of the mirror is slow enough to produce only negligible particle pairs from the field.  When one of the mirrors moves fast enough to create significant particles from the quantum field fluctuations we enter the related but separate realm of  {dynamical Casimir effect} \cite{Dodonov55} the full extent of which this paper will not delve into. However, for models with interaction functions that have a form falling under the categories of our functions $v$ and $u$, as shown in Sec.V.C, we can include those aspects and go to higher order in the displacement.

The alternative pathway we employ here, under the weak interaction assumption, is the functional perturbative method introduced in  Ref. \cite{HPZ93} which  works out for quantum Brownian motion \cite{HPZ92} (for the Schwinger-Keldysh closed-time-path and the Feynman-Vernon influence functional formalisms, see, e.g., \cite{CalHu08}, for its application to moving mirrors, see e.g., \cite{WuLee05,Fosco}) up to the second order in the interaction strength between the system (here, the mirror displacement) and its environment (here the quantum field)\footnote{In brief, for those readers familiar with \cite{HPZ93} which considered interactions of the type $f(x)q^k$ where $x$ and $q$ denote the system (mirror displacement) and bath (quantum field) variables respectively,  $k$ is the polynomial power in $q$  and $f(x)$ is an \textit{arbitrary} function of $x$, they would notice that $f(x)$ includes the power series expansion in $x$, the mirror displacement in the present context.}.  The recent paper Ref.\cite{ChoHu26} has extended to the third order in the system-bath interaction, and more relevantly for our purpose here, has included the momentum coupling terms, so one could deal with problems involving photon numbers,  common for QOM inquires. Our present work is a vivid application of the results obtain in \cite{ChoHu26} to QOM, as briefly discussed in its Sec. ID.  
After we have obtained our primary results in Sec. \ref{SecIF} and \ref{SecNonlinearLangevin} , we shall show how the other state-of-the-art models can be linked to the more general model presented here. The merits of this framework to cover a broader scope by means of a more powerful method will become clearer then. 

This paper is organized as follows: In Sec. \ref{SectQOMModels}, after commenting on the general aspects of the models used in QOM, we present our approach  which encompasses most if not all of the existing models. In Sec. \ref{SecIF} we introduce the Schwinger-Keldysh formalism for the calculation of the Feynman-Vernon influence action for the wall's degree of freedom up to the second order in the interaction strength between the wall and the cavity modes, where  the dissipation and noise kernels together with the potential renormalizations all show up. In Sec. \ref{SecNonlinearLangevin} we deduce the nonlinear Langevin equation of motion with the noise generated by the fluctuations of the cavity modes on the wall's motion. Sec. \ref{SecApplication} analyzes the different kinds of equations of motion according to the types of interaction models, focusing on the conditions under which nonlinearity and non-Markovian effects take place. Finally in Sec. \ref{SecConclu} we summarize our findings. In Appendix \ref{AppFunctionalDerivatives}, we give the details on the derivation of the functional derivatives required for the calculation of the influence action up to second order in the interaction strength, which is finally shown in Appendix \ref{App2ndIF}.

\section{QOM Hamiltonians: state-of-the-art and our model}\label{SectQOMModels}

{To provide a clear perspective in what we try to accomplish, in this section we give a brief overview of the important models used for QOM, in terms of both their compositions and their functionalities,  and where our model is situated in this landscape.}

We begin with the {well-known Hamiltonian accounting for radiation pressure phenomena as in} Ref.\cite{Macri2018}: The  {single} cavity field mode can be represented by an eigenmode of a quantum harmonic oscillator with  natural frequency $\omega$ whose quantum amplitudes are described by the canonical variables operators $\hat{q}$ and $\hat{p}_{q}$, associated with the lowering and raising operators $\hat{b}$ and $\hat{b}^{\dagger}$ in a Fock space representation. Attributing a mass ($m$) to the oscillator would correspond to a massive bosonic field \cite{ChoHu26}, whereby we have
\begin{equation}
\hat{b}=\frac{1}{\sqrt{2m\hbar\omega}}\left(m\omega\hat{q}+i\hat{p}_{q}\right)~~,~~\hat{b}^{\dagger}=\frac{1}{\sqrt{2m\hbar\omega}}\left(m\omega\hat{q}-i\hat{p}_{q}\right),
\label{bOperators}
\end{equation}
\begin{equation}
\hat{q}=\sqrt{\frac{\hbar}{2m\omega}}\left(\hat{b}+\hat{b}^{\dag}\right)~~,~~\hat{p}_{q}=i\sqrt{\frac{m\hbar\omega}{2}}\left(\hat{b}^{\dagger}-\hat{b}\right).
\end{equation}

The motion of the cavity's wall 
is described by the wall's displacement variable $Q$, whose quantum operator is $\hat{Q}$,  with canonical conjugate  momentum operator $\hat{P}$). In the presence of a trap the wall's motion will be subjected to an external potential $V(Q)$, such as the harmonic oscillator potential considered in \cite{Macri2018}. This setup is standard in the vast majority of models considering a single cavity mode. Differences among models come from different interaction Hamiltonians needed for the description of different phenomenology. For instance, in Ref.\cite{Macri2018} the movement of the wall while interacting with the field, or simply, the opto-mechanical interaction, takes on a more general form, written here with the notations of \cite{ChoHu26} as: 
\begin{equation}
\hat{H}_{\rm Int}=\hat{q}^{2}v(\hat{Q}),
\end{equation}
with the vertex function $v(\hat{Q})=\lambda C f(\hat{Q})$,  where  $C=m\omega G$ is the coupling constant, and $G$ is the same coupling parameter used in Ref.\cite{Macri2018}. This paper attempts to cover a broader range beyond small $Q$ by deriving stochastic dynamical equations with nonMarkovian backaction noises to capture effects from nonlinear displacements, therefore we opt for a fully general analytic function $f(Q)$ in the interaction. The special case  $f(\hat{Q})=\hat{Q}$ is the most commonly considered case in QOM, useful under the assumption of small displacements of the wall. However, the interaction between the cavity mode and the wall at the first order in the wall's displacement  {describing} the effect of radiation pressure is given by a simplified Hamiltonian. 

Processes involving  the creation or the annihilation of photon pairs are commonly associated with the dynamical Casimir effect, where the number of photons present in the cavity need not be conserved. For  small and/or slow displacements of the wall,  particle creation is insignificant. The terms in $\hat{H}_{\rm Int}$ accounting for these processes come from the terms $\hat{b}^{2}$ and $\hat{b}^{\dag 2}$ present in $\hat{q}^{2}$. Neglecting such terms, we have $\hat{H}_{\rm Int}\propto\hat{n}v(\hat{Q})$, accurately describing the phenomenology of radiation pressure in the lowest order. If we go back to a description in terms of position and momentum operators, as in Ref.\cite{ChoHu26}, we obtain the form of the interaction Hamiltonian conserving the number of photons as   $\hat{H}_{\rm Int}\propto(\omega^{2}\hat{q}^{2}+\hat{p}_{q}^{2}/m^{2})v(\hat{Q})$ plus some constant because of the non-commutativity of the canonical operators. Notice that equal interaction functions are found associated with $\omega^{2}\hat{q}^{2}$ and $\hat{p}_{q}^{2}/m^{2}$. However, this is not always the case.

{Along the existing pathway of carrying out a series expansion in the wall's displacement we recapitulate  the results of Ref.\cite{Ferreri2025} which seems to represent the state-of-the-art, namely, up to the third order in the wall's displacement along a direction $x$. Instead of treating a full 3D model,  we consider here a 1D model, which is simpler but sufficient enough to illustrate our approach mathematically and still useful physically. This  corresponds to taking the limit $k_{n_{\perp}}\rightarrow 0$, as shown by the authors of Ref.\cite{Ferreri2025}. For a field inside a rectangular cavity of length $L$ with one of the two walls moving in the $\pm x$ direction, using $\epsilon=\delta L/L$ to measure its relative displacement, these authors considered the following total Hamiltonian:
\begin{equation}
\hat{H}=\hat{H}_{0}+\epsilon\hat{H}_{1}+\epsilon^{2}\hat{H}_{2}+\epsilon^{3}\hat{H}_{3},
\label{HFull}
\end{equation}
where $\hat{H}_{0}=\sum_{n}\hbar\omega_{n}\hat{b}_{n}^{\dagger}\hat{b}_{n}+\hbar\Omega\hat{a}^{\dagger}\hat{a}$ is the free Hamiltonian of the field followed by that of the wall. Denoting the masses of the field oscillators by $m_{n}$ and $\Omega$ the oscillation frequency of the wall, the Hamiltonians $H_i$ corresponding to successive higher orders in $Q$, indicated by the subscript $i=1,2,3$, are given by
\begin{equation}
\hat{H}_{1}=-\sum_{n,m}(-1)^{n+m}\omega_{n}\omega_{m}\sqrt{m_{n}m_{m}}\hat{q}_{n}\hat{q}_{m}\frac{\hat{Q}}{Q_{\rm zp}},
\label{H1}
\end{equation}
\begin{equation}
\hat{H}_{2}=\sum_{n,m}(-1)^{n+m}\omega_{n}\omega_{m}\sqrt{m_{n}m_{m}}\hat{q}_{n}\hat{q}_{m}\left(\frac{\hat{Q}}{Q_{\rm zp}}\right)^{2},
\label{H2}
\end{equation}
\begin{equation}
\hat{H}_{3}=\sum_{n,m}(-1)^{n+m}\sqrt{m_{n}m_{m}}\left[\frac{k_{n}k_{m}L^{2}}{3}\frac{\hat{p}_{n}}{m_{n}}\frac{\hat{p}_{m}}{m_{m}}+\left(\frac{(k_{n}^{2}+k_{m}^{2})L^{2}}{6}-1\right)\omega_{n}\omega_{m}\hat{q}_{n}\hat{q}_{m}\right]\left(\frac{\hat{Q}}{Q_{\rm zp}}\right)^{3},
\label{H3}
\end{equation}
where $\omega_{n}=ck_{n}=n\pi c/L$ and $Q_{\rm zp}\equiv\sqrt{(\Delta\hat{Q})^{2}}=\sqrt{\hbar/(2M\Omega)}$ corresponds to the zero-point fluctuations of the wall's harmonic motion. Notice that mode mixing in a multi-mode cavity is obviously displayed here. This model allows to capture resonant interactions, the contribution of counterrotating terms, cross-Kerr frequency shifts and Kerr-like interactions at once. As a striking consequence, the functions accompanying the quadratic products of position and momentum operators result different.}


Unlike the approaches taken by the small number of engaging authors who have tackled this problem by performing a series expansion of the interaction Hamiltonian in orders of $Q$, we present an approach which can encompass arbitrary functional $f(\hat{Q})$ dependence on $Q$ -- the needed trade-off is weak couplings. Considering the last model, e.g., Ref.\cite{Ferreri2025}, as representative of the vast literature on quantum optomechanics,  we want to conceive a general model capable of keeping most of the features found there, namely, interactions with specific cavity modes (with photon number conservation or not) and cross terms between pairs of cavity modes. 
We write the actions by simply taking into account that the canonical momenta as $p_{n}=m_{n}\dot{q}_{n}$ and $P=M\dot{Q}$, while following the notation in Ref.\cite{ChoHu26}:
\begin{equation}
S=S_{0}+S_{\rm Int}~~,~~S_{0}[Q,\{q_{n}\}]=S_{0}^{\rm (f)}[\{q_{n}\}]+S_{0}^{\rm (w)}[Q],
\end{equation}
where the free field action splits into two actions for each subsystem with superscripts $(f)$ and $(w)$ denoting the field and the wall, respectively:
\begin{equation}
S_{0}^{\rm (f)}[\{q_{n}\}]=\int_{0}^{t}ds\sum_{n}\left[\frac{m_{n}}{2}\dot{q}_{n}^{2}-\frac{m_{n}}{2}\omega_{n}^{2}q_{n}^{2}\right]~~,~~S_{0}^{\rm (w)}[Q]=\int_{0}^{t}ds\left[\frac{M}{2}\dot{Q}^{2}-V(Q)\right],
\label{SFree}
\end{equation}
where $V(Q)$ is any arbitrary potential, albeit commonly assumed to be a harmonic oscillator potential  $V(Q)=M\Omega^{2}Q^{2}/2$.   The interaction action involves different vertex functions $v$ and $u$ for the positions and the momentum quadratic products:
\begin{equation}
S_{\rm int}\left[Q,\{q_{n}\}\right]=\int_{0}^{t}ds\sum_{n,m}\left[v_{nm}(Q)~\omega_{n}\omega_{m}q_{n}q_{m}+u_{nm}(Q)~\dot{q}_{n}\dot{q}_{m}\right].
\label{SintFull}
\end{equation}
As we will show more explicitly later, this interaction contains most of the common models in QOM as particular cases.

In the next sections, we perform an expansion in the small parameter $\lambda$ above, up to the second order, using functional perturbative methods developed in \cite{HPZ93}. In principle one can go to the third order in $\lambda^3$ following the results presented in \cite{ChoHu26}, but for our present purpose second order in $\lambda^2$ is sufficient to demonstrate the generality of this alternative approach.

\section{Influence functional of the quantum field on the wall's motion}\label{SecIF}
We now show our approach to problems in QOM following Ref.\cite{ChoHu26} in two steps: 1) Derivation of the influence functional which captures the influence of the quantum field environment, in terms of the quantum noises therein, on the motion of the wall; 2) Derivation of the Langevin equation with nonMarkovian noises as source from taking the functional variation of the influence functional. We do 1) in this section and 2) in the next.  

Using the closed-time-path formalism, the influence functional is given by:
\begin{eqnarray}
F\left[Q_{+},Q_{-}\right]&=&e^{\frac{i}{\hbar}S_{\rm IF}\left[Q_{+},Q_{-}\right]}\\
&=&\int_{\rm CTP}\prod_{n}Dq_{n+}Dq_{n-}~e^{\frac{i}{\hbar}\left(S_{0}^{\rm (f)}\left[\{q_{n+}\}\right]+S_{\rm int}\left[Q_{+},\{q_{n+}\}\right]-S_{0}^{\rm (f)}\left[\{q_{n-}\}\right]-S_{\rm int}\left[Q_{-},\{q_{n-}\}\right]\right)},\nonumber
\end{eqnarray}
where $S_{\rm IF}[Q_{+},Q_{-}]$ is the influence action. In the functional perturbative approach we consider the exponentials in the interaction action $S_{\rm int}$ for each variable, expand and replace them by the functional derivatives with respect to a pair of sources $J_{n \pm}$ for each cavity mode $n$ in the generating functional, arriving at the correlation functions:
\begin{equation}
F\left[Q_{+},Q_{-}\right]=\left.e^{\frac{i}{\hbar}\left(S_{\rm int}\left[Q_{+},\left\{\frac{\hbar}{i}\frac{\delta}{\delta J_{n+}}\right\}\right]-S_{\rm int}\left[Q_{-},\left\{-\frac{\hbar}{i}\frac{\delta}{\delta J_{n-}}\right\}\right]\right)}Z\left[\left\{J_{n+}\right\},\left\{J_{n-}\right\}\right]\right|_{J_{n+}=J_{n-}=0},
\label{IFIntDerivatives}
\end{equation}
where $Z[\{J_{n+}\},\{J_{n-}\}]$ stands for the generating functional for the cavity modes' correlations, given by:
\begin{eqnarray}
&&Z\left[\left\{J_{n+}\right\},\left\{J_{n-}\right\}\right]=\\
&&=\int_{\rm CTP}\prod_{n}Dq_{n+}Dq_{n-}~e^{\frac{i}{\hbar}\left(S_{0}^{\rm (f)}\left[\{q_{n+}\}\right]+\int_{0}^{t}ds\sum_{n}J_{n+}(s)q_{n+}(s)-S_{0}^{\rm (f)}\left[\{q_{n-}\}\right]-\int_{0}^{t}ds\sum_{n}J_{n-}(s)q_{n-}(s)\right)}.\nonumber
\end{eqnarray}

Notice that we are including a pair of source $\{J_{n+},J_{n-}\}$ per cavity mode. This is required in order to obtain cross terms between the cavity modes, as given in the interaction model as can be observed in Eq.(\ref{SintFull}). On this point, the present case is a generalization of that treated in Ref.\cite{ChoHu26}.

Given that $S_{0}^{\rm (f)}[\{q_{n}\}]=\sum_{n}S_{0}^{\rm (f)}[q_{n}]$, the generating functional can be split  as a product of the generating functional for each oscillator:
\begin{equation}
Z\left[\left\{J_{n+}\right\},\left\{J_{n-}\right\}\right]=\prod_{n}Z^{(1)}\left[J_{n+},J_{n-}\right],
\label{FullZProduct}
\end{equation}
where the generating functional of each oscillator reads:
\begin{eqnarray}
Z^{(1)}\left[J_{n+},J_{n-}\right]&=&\int_{\rm CTP}Dq_{n+}Dq_{n-}~e^{\frac{i}{\hbar}\left(S_{0}^{\rm (f)}\left[q_{n+}\right]+\int_{0}^{t}ds~J_{n+}(s)q_{n+}(s)-S_{0}^{\rm (f)}\left[q_{n-}\right]-\int_{0}^{t}ds~J_{n-}(s)q_{n-}(s)\right)}\\
&=&{\rm Exp}\Bigg(\frac{i}{2\hbar^{2}}\int_{0}^{t}ds\int_{0}^{t}ds'\left[J_{n+}(s)G_{n++}(s,s')J_{n+}(s')-J_{n+}(s)G_{n+-}(s,s')J_{n-}(s')\right.\nonumber\\
&&-\left.J_{n-}(s)G_{n-+}(s,s')J_{n+}(s')+J_{n-}(s)G_{n--}(s,s')J_{n-}(s')\right]\Bigg),\nonumber
\end{eqnarray}
The propagators are the same as in Ref.\cite{ChoHu26}:
\begin{equation}
G_{n++}(s,s')=\hbar\left[-\mu_{n}(s-s'){\rm sgn}(s-s')+i\nu_{n}(s-s')\right],
\label{Gn++}
\end{equation}
\begin{equation}
G_{n+-}(s,s')=\hbar\left[\mu_{n}(s-s')+i\nu_{n}(s-s')\right],
\label{Gn+-}
\end{equation}
\begin{equation}
G_{n-+}(s,s')=\hbar\left[-\mu_{n}(s-s')+i\nu_{n}(s-s')\right],
\label{Gn-+}
\end{equation}
\begin{equation}
G_{n--}(s,s')=\hbar\left[\mu_{n}(s-s'){\rm sgn}(s-s')+i\nu_{n}(s-s')\right],
\label{Gn--}
\end{equation}
with:
\begin{equation}
\mu_{n}(s-s')=-\frac{1}{2m_{n}\omega_{n}}\sin\left[\omega_{n}(s-s')\right]~~,~~\nu_{n}(s-s')=\frac{1}{2m_{n}\omega_{n}}\cos\left[\omega_{n}(s-s')\right].
\label{munukernels}
\end{equation}

To facilitate a perturbative treatment for weak opto-mechanical coupling we have included a dimensionless perturbation parameter $\lambda$ associated with the total interaction action as done in Ref.\cite{ChoHu26}. Instead of giving each term as a sum over connected diagrams, we proceed a bit differently by considering that the influence action can be written as:
\begin{equation}
S_{\rm IF}\left[Q_{+},Q_{-}\right]=-i\hbar~{\rm ln}\left(1+\sum_{j=1}^{+\infty}\left[\frac{i\lambda}{\hbar}\right]^{j}F^{(j)}\left[Q_{+},Q_{-}\right]\right),
\label{InfluenceActionFull}
\end{equation}
where each $F^{(j)}$ is the contribution to the influence action at a given power of $\lambda$ which are given by:
%
%
%
%
\begin{eqnarray}
&&F^{(j)}\left[Q_{+},Q_{-}\right]=\\
&&=\frac{1}{j!}\left.\left(S_{\rm int}\left[Q_{+},\left\{\frac{\hbar}{i}\frac{\delta}{\delta J_{n+}}\right\}\right]-S_{\rm int}\left[Q_{-},\left\{-\frac{\hbar}{i}\frac{\delta}{\delta J_{n-}}\right\}\right]\right)^{j}Z\left[\left\{J_{n+}\right\},\left\{J_{n-}\right\}\right]\right|_{J_{n+}=J_{n-}=0},\nonumber
\end{eqnarray}
instead of considering each just as the connected diagrams as in Ref.\cite{ChoHu26}. Nevertheless, the connection between the two approaches comes in the expansion of the influence action of Eq.(\ref{InfluenceActionFull}) in powers of $\lambda$. We show the expression up to the third order in $\lambda$ below:
\begin{eqnarray}
S_{\rm IF}\left[Q_{+},Q_{-}\right]&=&\sum_{j=1}^{+\infty}\lambda^{j}S_{\rm IF}^{(j)}\left[Q_{+},Q_{-}\right]\label{SIFExpansion}\\
&\approx&\lambda F^{(1)}\left[Q_{+},Q_{-}\right]-\frac{i\lambda^{2}}{2\hbar}\left[\left(F^{(1)}\left[Q_{+},Q_{-}\right]\right)^{2}-2F^{(2)}\left[Q_{+},Q_{-}\right]\right]\nonumber\\
&&-\frac{\lambda^{3}}{3\hbar^{2}}\left[\left(F^{(1)}\left[Q_{+},Q_{-}\right]\right)^{3}-3F^{(1)}\left[Q_{+},Q_{-}\right]F^{(2)}\left[Q_{+},Q_{-}\right]+3F^{(3)}\left[Q_{+},Q_{-}\right]\right]+...\nonumber
\end{eqnarray}
Notice that each term includes contributions from  previous orders, and as such, it should agree with the approach taken in Ref.\cite{ChoHu26} of connected diagrams.

We wish to calculate $S_{\rm IF}^{(j)}$ for $j=1,2,3$. For $j=1$, we have:
\begin{eqnarray}
&&S_{\rm IF}^{(1)}\left[Q_{+},Q_{-}\right]=\\
&&=-\hbar^{2}\int_{0}^{t}ds\sum_{n,m}\left(v_{nm}(Q_{+})\omega_{n}\omega_{m}~\frac{\delta}{\delta J_{n+}(s)}\frac{\delta}{\delta J_{m+}(s)}+u_{nm}(Q_{+})\left.\partial^{2}_{\tau\tau'}\frac{\delta}{\delta J_{n+}(\tau)}\frac{\delta}{\delta J_{m+}(\tau')}\right|_{\tau=\tau'=s}\right.\nonumber\\
&&-\left.v_{nm}(Q_{-})\omega_{n}\omega_{m}~\frac{\delta}{\delta J_{n-}(s)}\frac{\delta}{\delta J_{m-}(s)}-u_{nm}(Q_{-})\left.\partial^{2}_{\tau\tau'}\frac{\delta}{\delta J_{n-}(\tau)}\frac{\delta}{\delta J_{m-}(\tau')}\right|_{\tau=\tau'=s}\right)\nonumber\\
&&\times\left.Z\left[\left\{J_{n+}\right\},\left\{J_{n-}\right\}\right]\right|_{J_{n+}=J_{n-}=0}.\nonumber
\end{eqnarray}

Following Ref.\cite{ChoHu26}, we have that the time derivatives on the terms associated with the functions $u_{mn}$ directly apply on the Green functions of the environmental degrees of freedom resulting from the functional derivation. Considering $G_{n\pm\pm}$ are symmetric in the exchange $s\leftrightarrow s'$, the functional derivatives give:
\begin{equation}
\left.\frac{\delta^{2}Z}{\delta J_{n\pm}(\tau)\delta J_{m\pm}(\tau')}\right|_{J_{k+}=J_{k-}=0}=\delta_{nm}\frac{i}{\hbar^{2}}G_{n\pm\pm}(\tau,\tau'),
\end{equation}
where the $\delta_{nm}$ enters because there are no cross  terms between the sources $\{J_{n\pm}\}$ in Eq.(\ref{FullZProduct}). Then, we have:
\begin{eqnarray}
S_{\rm IF}^{(1)}\left[Q_{+},Q_{-}\right]&=&-i\int_{0}^{t}ds\sum_{n}\left[v_{nn}(Q_{+})\omega_{n}^{2}G_{n++}(0)
+u_{nn}(Q_{+})\partial^{2}_{12}G_{n++}(0)
-v_{nn}(Q_{-})\omega_{n}^{2}G_{n--}(0)
\right.\nonumber\\
&&-\left.u_{nn}(Q_{-})\partial^{2}_{12}G_{n--}(0)
\right].\nonumber
\label{SIF1G}
\end{eqnarray}

Evaluating $S_{\rm IF}^{(1)}$ involves evaluations of $G_{n\pm\pm}$ and its crossed second derivatives for $\tau=\tau'=s$, which give:
\begin{equation}
G_{n\pm\pm}(0)
=iq_{n,{\rm zp}}^{2}~~,~~\partial^{2}_{12}G_{n\pm\pm}(0)
=iq_{n,{\rm zp}}^{2}\omega_{n}\left[\pm i2\delta(0)+\omega_{n}\right],
\end{equation}
where $q_{n,{\rm zp}}=\sqrt{\hbar/(2m_{n}\omega_{n})}$ is the zero-point fluctuations of the $n-$th cavity mode, while the term coming with $\delta(0)$ corresponds to the infinite renormalization already found in Ref.\cite{ChoHu26}. Then, to the first order in the coupling constant $\lambda$, it reads:
\begin{equation}
S_{\rm IF}^{(1)}\left[Q_{+},Q_{-}\right]
=\int_{0}^{t}ds\left(-\sum_{n}\delta V_{n+}^{(1)}(Q_{+})\right)-\int_{0}^{t}ds\left(-\sum_{n}\delta V_{n-}^{(1)}(Q_{-})\right),
\label{SIF1Potentials}
\end{equation}
where each potential renormalization read:
\begin{equation}
\delta V_{n\pm}^{(1)}(Q_{\pm})=-q_{n,{\rm zp}}^{2}\omega_{n}\left(\pm 2i\delta(0)u_{nn}\left(Q_{\pm}\right)+\omega_{n}\left[v_{nn}\left(Q_{\pm}\right)+u_{nn}\left(Q_{\pm}\right)\right]\right).
\label{dVn1}
\end{equation}

Now, to calculate $S_{\rm IF}^{(2)}$, the second order influence action, $F^{(2)}$ is required:
\begin{equation}
F^{(2)}\left[Q_{+},Q_{-}\right]=\frac{1}{2}\left.\left(S_{\rm int}\left[Q_{+},\left\{\frac{\hbar}{i}\frac{\delta}{\delta J_{n+}}\right\}\right]-S_{\rm int}\left[Q_{-},\left\{-\frac{\hbar}{i}\frac{\delta}{\delta J_{n-}}\right\}\right]\right)^{2}Z\left[\left\{J_{n+}\right\},\left\{J_{n-}\right\}\right]\right|_{J_{n+}=J_{n-}=0}.
\label{SIF2}
\end{equation}

The full calculation of the latter is shown in App.\ref{App2ndIF}, where we directly go to the second order correction appearing in Eq.(\ref{SIFExpansion}), in such a way that we obtain:
\begin{eqnarray}
S_{\rm IF}^{(2)}\left[Q_{+},Q_{-}\right]&=&\int_{0}^{t}ds\left(-\sum_{a,b}\delta V_{ab+}^{(2)}(Q_{+})\right)-\int_{0}^{t}ds\left(-\sum_{a,b}\delta V_{ab-}^{(2)}(Q_{-})\right)\label{SIF1MinusSIF2}\\
&&+\int_{0}^{t}ds\int_{0}^{t}ds'\sum_{a,b}\left(-\left[\Delta v_{ab}(s)-\Delta u_{ab}(s)\right]2\mathcal{D}_{ab}(s,s')\left[\Sigma v_{ab}(s')-\Sigma u_{ab}(s')\right]\right.\nonumber\\
&&+\left.\left[\Delta v_{ab}(s)-\Delta u_{ab}(s)\right]\frac{i}{2}\mathcal{N}_{ab}(s,s')\left[\Delta v_{ab}(s')-\Delta u_{ab}(s')\right]\right),\nonumber
\end{eqnarray}
where the correction to the potentials is:
\begin{equation}
\delta V_{ab\pm}^{(2)}(Q_{\pm})=\frac{1}{\hbar}\omega_{a}\omega_{b}q_{a,{\rm zp}}^{2}q_{b,{\rm zp}}^{2}\left[\omega_{a}+\omega_{b}\pm 4i\delta(0)\right]u_{ab}(Q_{\pm})u_{ab}(Q_{\pm}),
\label{dVab2}
\end{equation}
which again includes finite and divergent terms which require renormalization, but notice they involve only  $u_{ab}$ and not $v_{ab}$, while, on the other hand, the dissipation and noise kernels reads:
\begin{equation}
\mathcal{D}_{ab}(s,s')=\theta(s-s')\frac{2}{\hbar}\omega_{a}^{2}\omega_{b}^{2}q_{a,{\rm zp}}^{2}q_{b,{\rm zp}}^{2}\sin\left[\left(\omega_{a}+\omega_{b}\right)(s-s')\right],
\label{Dissab}
\end{equation}
\begin{equation}
\mathcal{N}_{ab}(s,s')=\frac{2}{\hbar}\omega_{a}^{2}\omega_{b}^{2}q_{a,{\rm zp}}^{2}q_{b,{\rm zp}}^{2}\cos\left[\left(\omega_{a}+\omega_{b}\right)(s-s')\right].
\label{Noiab}
\end{equation}
Both kernels are real, the dissipation kernel being a causal function while the  noise kernel is symmetric under the exchange $s\leftrightarrow s'$.

All told, Eqs.(\ref{SIF1Potentials}) and (\ref{SIF1MinusSIF2}) can be inserted into Eq.(\ref{SIFExpansion}) to obtain the influence action $S_{\rm IF}$ up to second order in $\lambda$.

\section{nonlinear Langevin equation with non-Markovian backaction noises}\label{SecNonlinearLangevin}

Having obtained the influence action up to second order, the coarse-grained effective action is written as $S_{\rm CGEF}[Q_{+},Q_{-}]=S_{0}^{\rm (w)}[Q_{+}]-S_{0}^{\rm (w)}[Q_{-}]+S_{\rm IF}[Q_{+},Q_{-}]$, in such a way that:
\begin{eqnarray}
S_{\rm CGEF}\left[Q_{+},Q_{-}\right]&\approx& S_{0}^{\rm (w)}[Q_{+}]-S_{0}^{\rm (w)}[Q_{-}]-\int_{0}^{t}ds~\delta V_{+}\left(Q_{+}\right)+\int_{0}^{t}ds~\delta V_{-}\left(Q_{-}\right)\\
&&+\int_{0}^{t}ds\int_{0}^{t}ds'\sum_{a,b}\left(-\left[\Delta v_{ab}(s)-\Delta u_{ab}(s)\right]2\mathcal{D}_{ab}(s,s')\left[\Sigma v_{ab}(s')-\Sigma u_{ab}(s')\right]\right.\nonumber\\
&&+\left.\left[\Delta v_{ab}(s)-\Delta u_{ab}(s)\right]\frac{i}{2}\mathcal{N}_{ab}(s,s')\left[\Delta v_{ab}(s')-\Delta u_{ab}(s')\right]\right),\nonumber
\end{eqnarray}
where the dissipation and noise kernels are given by Eqs.(\ref{Dissab}) and (\ref{Noiab}), and the renormalization of the potentials are:
\begin{equation}
\delta V_{\pm}\left(Q_{\pm}\right)=\sum_{a}\delta V_{a\pm}^{(1)}\left(Q_{\pm}\right)+\sum_{a,b}\delta V_{ab\pm}^{(2)}\left(Q_{\pm}\right),
\label{FullPotentialRenormalization}
\end{equation}
given Eqs.(\ref{dVn1}) and (\ref{dVab2}).

At this point, by observing the quadratic form of the terms associated with the noise kernels, it is useful to consider them as the effects of stochastic forces. In order to define the stochastic forces in a suitable way, notice that the symmetry of the functions $v_{ab}$ and $u_{ab}$ and the noise kernels $\mathcal{N}_{ab}$ involves that the sums over $a$ and $b$ can be written as $\sum_{a,b}=\sum_{a}\delta_{ab}+2\sum_{a<b}$. Re-writing all these terms accordingly, we take a step further to match every combination $\{a,b\}$ into a single label $\alpha$, simplifying the notation. For instance, taking $N$ cavity modes, $1<\alpha<N(N+1)/2$ according to the assignment $\{a,b\}\rightarrow\{\alpha\}$ under the condition $b\geqslant a$, we have:
\begin{equation}
\sum_{a,b}\left[\Delta v_{ab}(s)-\Delta u_{ab}(s)\right]\frac{i}{2}\mathcal{N}_{ab}(s,s')\left[\Delta v_{ab}(s')-\Delta u_{ab}(s')\right]=\sum_{\alpha=1}^{\frac{N(N+1)}{2}}\Delta f_{\alpha}(s)\frac{i}{2}\widetilde{\mathcal{N}}_{\alpha}(s,s')\Delta f_{\alpha}(s'),
\label{Fdefinition}
\end{equation}
where $f_{\alpha}=v_{\alpha}-u_{\alpha}$ and $\widetilde{\mathcal{N}}_{\alpha}=C_{\alpha}\mathcal{N}_{\alpha}$, where $C_{\alpha}=1$ if $\alpha$ corresponds to a pair $\{a,b\}$ such that $a=b$, while $C_{\alpha}=2$ for $\alpha$ such that $a\neq b$.

Given this new form, following Ref.\cite{ChoHu26}, it is possible to define a stochastic force for each of the $\alpha$ according to:
\begin{eqnarray}
e^{-\frac{1}{2\hbar}\int_{0}^{t}ds\int_{0}^{t}ds'\sum_{\alpha=1}^{N(N+1)/2}\Delta f_{\alpha}(s)\widetilde{\mathcal{N}}_{\alpha}(s,s')\Delta f_{\alpha}(s')}&=&\prod_{\alpha=1}^{\frac{N(N+1)}{2}}e^{-\frac{1}{2\hbar}\int_{0}^{t}ds\int_{0}^{t}ds'\Delta f_{\alpha}(s)\widetilde{\mathcal{N}}_{\alpha}(s,s')\Delta f_{\alpha}(s')}\\
&=&\prod_{\alpha=1}^{\frac{N(N+1)}{2}}\int D\xi_{\alpha}~\mathcal{P}_{\alpha}\left[\xi_{\alpha}\right]~e^{\frac{i}{\hbar}\int_{0}^{t}ds\Delta f_{\alpha}(s)\xi_{\alpha}(s)},\nonumber
\end{eqnarray}
where $\mathcal{P}_{\alpha}[\xi_{\alpha}]$ is the probability density for the stochastic force $\xi_{\alpha}(s)$. As the result of each integral is a Gaussian, the probability densities are also Gaussian. They are sufficiently characterized by their first two moments:
\begin{equation}
\left\langle\xi_{\alpha}(s)\right\rangle=0~~,~~\left\langle\xi_{\alpha}(s)\xi_{\beta}(s')\right\rangle=\delta_{\alpha\beta}~\hbar\widetilde{\mathcal{N}}_{\alpha}(s,s'),
\end{equation}
where the Kronecker delta implies that the different stochastic forces are uncorrelated each other. In this sense, each probability density can be written as:
\begin{equation}
\mathcal{P}_{\alpha}\left[\xi_{\alpha}\right]=\mathcal{K}_{\alpha}~e^{-\frac{1}{2\hbar}\int_{0}^{t}ds\int_{0}^{t}ds'\xi_{\alpha}(s)\left[\widetilde{\mathcal{N}}_{\alpha}(s,s')\right]^{-1}\xi_{\alpha}(s')},
\end{equation}
where the normalization constant reads:
\begin{equation}
\mathcal{K}_{\alpha}=\left(\int D\xi_{\alpha}~e^{-\frac{1}{2\hbar}\int_{0}^{t}ds\int_{0}^{t}ds'\xi_{\alpha}(s)\left[\widetilde{\mathcal{N}}_{\alpha}(s,s')\right]^{-1}\xi_{\alpha}(s')}\right)^{-1},
\end{equation}
having $[\widetilde{\mathcal{N}}_{\alpha}(s,s')]^{-1}$ as the inverse bifunction defined by:
\begin{equation}
\int_{0}^{t}ds''\widetilde{\mathcal{N}}_{\alpha}(s,s'')\left[\widetilde{\mathcal{N}}_{\alpha}(s'',s')\right]^{-1}=\delta(s-s').
\end{equation}

Given the latter, it is possible to define a stochastic effective action including these stochastic forces, such that:
%
%
\begin{eqnarray}
\Gamma\left[Q_{+},Q_{-}\right]&=&\int_{0}^{t}ds\left[\frac{M}{2}\dot{Q}_{+}^{2}-\widetilde{V}_{+}\left(Q_{+}\right)\right]-\int_{0}^{t}ds\left[\frac{M}{2}\dot{Q}_{-}^{2}-\widetilde{V}_{-}\left(Q_{-}\right)\right]\\
&&-2\int_{0}^{t}ds\int_{0}^{s}ds'\sum_{\alpha=1}^{\frac{N(N+1)}{2}}\Delta f_{\alpha}(s)\widetilde{\gamma}_{\alpha}(s,s')\Sigma\dot{f}_{\alpha}(s')+\int_{0}^{t}ds\sum_{\alpha=1}^{\frac{N(N+1)}{2}}\Delta f_{\alpha}(s)\xi_{\alpha}(s),\nonumber
\end{eqnarray}
where $\widetilde{V}_{\pm}(Q_{\pm})=V(Q_{\pm})+\delta V_{\pm}(Q_{\pm})$ is the total renormalized potential and   $\widetilde{\mathcal{D}}_{\alpha} (s,s')=d\widetilde{\gamma}_{\alpha}(s,s')/ds$ is the damping kernel. 

From this stochastic effective action, we can derive an stochastic equation of motion for the wall's degree of freedom $Q$ similarly to what is done in Ref.\cite{ChoHu26}:
\begin{equation}
M\ddot{Q}+\widetilde{V}'_{+}(Q)+\int_{0}^{s}d\tau\sum_{\alpha=1}^{\frac{N(N+1)}{2}}\widetilde{\gamma}_{\alpha}(s,\tau)f'_{\alpha}\left(Q(s)\right)f'_{\alpha}\left(Q(\tau)\right)\dot{Q}(\tau)=\sum_{\alpha=1}^{\frac{N(N+1)}{2}}f'_{\alpha}\left(Q(s)\right)\xi_{\alpha}(s).
\label{NonlinearLangevinEquation}
\end{equation}

This is a nonlinear Langevin equation in the form of a single degree of freedom $Q$ nonlinearly coupled to $N(N+1)/2$ uncorrelated baths in our general model. We can understand this as a combination of i) $N$ of these noises arising from each individual cavity mode, and ii)  ($N(N-1)/2$) noises arising from the cross terms involving cavity modes interactions. 
When there is no interaction between the cavity modes component ii) will be absent, the sum will then run from 1 to $N$. Moreover, for interaction functions of the form $f_{\alpha}(Q)=\lambda C_{\alpha}f(Q)$, i.e., the same dependence on $Q$ for all the cavity modes, the nonlinear Langevin equation reads:
\begin{equation}
M\ddot{Q}+\widetilde{V}'_{+}(Q)+\int_{0}^{s}d\tau\widetilde{\gamma}(s,\tau)f'\left(Q(s)\right)f'\left(Q(\tau)\right)\dot{Q}(\tau)= f'\left(Q(s)\right)\xi(s).
\label{NonlinearChoHuEquation}
\end{equation}
with the total damping kernel and stochastic source given by:
\begin{equation}
\widetilde{\gamma}(s,\tau)=\lambda^{2}\sum_{\alpha=1}^{N}C_{\alpha}^{2}\widetilde{\gamma}_{\alpha}(s,\tau)~~,~~\xi(s)=\lambda\sum_{\alpha=1}^{N}C_{\alpha}\xi_{\alpha}(s),
\end{equation}
which reduces to an equation obtainable from the calculations in Ref.\cite{ChoHu26} by keeping only up to  the second order instead to the third order as its authors do.

In conclusion, Eq.(\ref{NonlinearLangevinEquation}) is the sought-after nonlinear Langevin equation with non-Markovian noises arising from the quantum field, including radiation pressure and fluctuations backaction. It works only for a weakly-coupled quantum opto-mechanical system up to the second order in the coupling constant $\lambda$, but is valid for any arbitrary functional dependence of the wall's displacement $Q$, not just restricted to power laws $Q^n$. In the next section we try to connect our general results to several popular models studied in the literature. 

\section{Relating to generic properties of popular optomechanical models}\label{SecApplication}

For the general model described by Eqs.(\ref{SFree}) and (\ref{SintFull}) we have derived the nonlinear Langevin equation  Eq.(\ref{NonlinearLangevinEquation}) up to the second order in the opto-mechanical interaction strength.  Under different approximations, in different regimes and describing different phenomenology, there is a plethora of optomechanical models found in the literature with different kinds of master equations, Langevin equations and stochastic Schr\"odinger equations. In this section, we shall try to connect the model studied here and the  results we obtained with some of the popular models in QOM and their Langevin equations and discuss their characteristic features.

The kind of effective equation of motion directly depends on the interaction functions  $f_{\alpha} \equiv v_{\alpha}-u_{\alpha}$ defined {in Eq.(\ref{Fdefinition}), weighting the asymmetry between $v_{\alpha}$ and $u_{\alpha}$. It also enters in the potential renormalization of Eq.(\ref{FullPotentialRenormalization}). In our formalism the terms involving a given cavity mode ($\alpha\leftrightarrow\{n,n\}$) and the terms involving two cavity modes ($\alpha\leftrightarrow\{n,m\}$ with $n\neq m$) appear on equal footing in the equation. Because of this we can analyze different cases of $f_{\alpha}$ independently,  whether it corresponds to a  single mode or two modes. For a general interaction hamiltonian, we need to analyze different terms each and then combine their contributions to obtain the corresponding equation of motion.

\subsection{The case $f_{\alpha}=0$ and the radiation pressure models}\label{SectF0}

The simplest contribution corresponds to interaction terms with  an interaction function $f_{\alpha}=0$. This implies $v_{nm}=u_{nm}$. For this case, there are no associated dissipation nor noise terms contributing to the equation. But the renormalization of the potential remains in the simplified versions:
\begin{equation}
\delta V_{n+}^{(1)}(Q)=-2q_{n,{\rm zp}}^{2}\omega_{n}\left(i\delta(0)+\omega_{n}\right)v_{nn}\left(Q\right),
\end{equation}
\begin{equation}
\delta V_{nm+}^{(2)}(Q)=\frac{1}{\hbar}\omega_{n}\omega_{m}q_{n,{\rm zp}}^{2}q_{m,{\rm zp}}^{2}\left[\omega_{n}+\omega_{m}+ 4i\delta(0)\right]v_{nm}^{2}(Q).
\end{equation}

Given this, the effective equation of motion reads:
\begin{eqnarray}
M\ddot{Q}+M\Omega^{2}Q&-&\sum_{n}2\omega_{n}q_{n,{\rm zp}}^{2}\left[\omega_{n}+i\delta(0)\right]v'_{nn}\left(Q\right)\label{EqMultiEqual}\\
&+&\sum_{n,m}\frac{2}{\hbar}\omega_{n}\omega_{m}q_{n,{\rm zp}}^{2}q_{m,{\rm zp}}^{2}\left[\omega_{n}+\omega_{m}+ 4i\delta(0)\right]v_{nm}(Q)v'_{nm}(Q)=0.\nonumber
\end{eqnarray}

This type of equation strongly depends on the dependence of the functions $v_{nm}$. Nonlinear equations are possible for nonlinear dependence of $v_{nm}$. Notice that although the equations could be nonlinear, this case leads only to equations of motion local  in time.

This simplest case corresponds to the most popular QOM model where the interaction between the cavity modes and the wall goes by radiation pressure. This is obtained when considering small amplitude motion of the wall and neglecting the terms for the creation and annihilation of photon pairs associated with  dynamical Casimir effects. Under these conditions, the interaction hamiltonian is proportional to the product $\hat{n}_{a}v_{a}(\hat{Q})$, which implies $v_{nn}=u_{nn}$ for each cavity mode while $v_{nm}=0=u_{nm}$ when $n\neq m$, namely, no interaction between modes.

This type of effective equations is obtained for models studied, e.g., in Ref.\cite{Macri2018} with exact solutions obtained numerically.  For this case,  $v_{nm}(Q)=u_{nm}(Q)=\delta_{n,1}\delta_{nm}\lambda CQ$, with linear dependence on $Q$. And, because $f_{\alpha}=0$ there are no dissipation or noise terms.  The only effects correspond to potential renormalizations $\delta V_{+}(Q)$ and to a constant applied force term arising from the commutation relations between $\hat{q}$ and $\hat{p}$ when writing the number operator $\hat{n}_{q}$. 
Under these conditions, the effective equation of motion up to second order of the interaction strength $\lambda$ results:
\begin{equation}
M\ddot{Q}+M\Omega_{\rm SE}^{2}Q=\mathcal{F}_{\rm SE}~~,~~\Omega_{\rm SE}^{2}=\Omega^{2}+\frac{2}{M}\omega^{2}q_{\rm zp}^{4}\lambda^{2}C^{2}[2i\delta(0)+\omega]~~,~~\mathcal{F}_{\rm SE}=2\omega q_{\rm zp}^{2}\lambda C[i\delta(0)+2\omega],
\label{EqSingleEqual}
\end{equation}
It has the form of the equation of motion for a simple harmonic oscillator of frequency $\Omega_{\rm SE}$ under the influence  of a constant force $\mathcal{F}_{\rm SE}$. The second term $2\omega$ within the square brackets of $\mathcal{F}_{\rm SE}$ comes from the aforementioned contribution of the commutation relations between $\hat{q}$ and $\hat{p}$.

Similar nature can be found in models discussed in Ref.\cite{BarchielliVacchini2015} which presents an extensive theoretical analysis of quantum Langevin equations in optomechanics. However, their models contain additional features like dissipation and noise from thermal reservoirs and  cavity photon pumping by an applied laser.  
These features can be easily added to our modeling. E.g.,  thermal reservoirs can be included through linearly coupled sets of harmonic oscillators in the Caldeira-Leggett model \cite{CalLeg83}, see, e.g.,  Ref.\cite{CalHu08}. In addition to these aspects, also in this case, the interaction with the cavity mode up to second order results in an equation of the form of Eq.(\ref{EqSingleEqual}) by suitably fixing $C$.

A comparison is also possible with the models employed in Ref.\cite{Mey08}. Either the so-called `linear' or `quadratic' models present an optomechanical interaction of the form $\hat{n}_{a}v_{a}(\hat{Q})$. Similarly it happens for the more exotic model of Ref.\cite{Vitali2025}, although it includes two movable walls. Additional influences from thermal baths and pumpings must be included. Besides this, the set-up is given by a cavity of fixed plates with a third movable one in-between. The scenario is described by a Hamiltonian involving two types of cavity modes, separated into even and odd, depending on how they change at each side of the movable plate. Nevertheless, since each mode fulfills the condition $v_{nn}=u_{nn}$, so this simple case in our model is still applicable, with two cavity modes coupled to the movable plate. From Eq.(\ref{EqMultiEqual}) for the multimode cases, we choose two cavity modes ($1$ and $2$) influencing the wall but without direct interaction among them. Setting $v_{2}(Q)=-v_{1}(Q)=-v(Q)$ the equation reads:
\begin{equation}
M\ddot{Q}+M\Omega^{2}Q+2v'(Q)\left[\mathcal{S}+2v(Q)\mathcal{T}\right]=0.
\end{equation}
with the factors $\mathcal{S}$ and $\mathcal{T}$ given by:
\begin{equation}
\mathcal{S}=\sum_{n=1}^{2}(-1)^{n}\omega_{n}q_{n,{\rm zp}}^{2}\left[\omega_{n}+i\delta(0)\right]~~,~~\mathcal{T}=\frac{1}{\hbar}\sum_{n=1}^{2}\omega_{n}^{2}q_{n,{\rm zp}}^{4}\left[\omega_{n}+2i\delta(0)\right].
\end{equation}

Given this equation, for the `linear' model of Ref.\cite{Mey08}, the equation of motion corresponds to the one of a simple harmonic oscillator with frequency renormalization and constant forcing. 
More interesting dynamics arise for the `quadratic' model of Ref.\cite{Mey08}, with the equation for a Duffing oscillator without dissipation nor periodic driving force. In this situation, we have $v(Q)=\lambda M\mathcal{A}Q^{2}$, with suitable units for $\mathcal{A}$, in such a way that the equation of motion is $\ddot{Q}+(\Omega^{2}+4\lambda\mathcal{A}\mathcal{S})Q+4\lambda^{2}\mathcal{A}^{2}\mathcal{T}MQ^{3}=0.$. From this it immediately turns out that $\dot{Q}^{2}/2+(\Omega^{2}+4\lambda\mathcal{A}\mathcal{S})Q^{2}/2+\lambda^{2}\mathcal{A}^{2}\mathcal{T}MQ^{4}=\mathcal{H}$, with $\mathcal{H}$ a constant, meaning that this quantity is a constant of motion. Moreover, given that $\lambda^{2}\mathcal{A}^{2}\mathcal{T}M>0$, whenever $\Omega^{2}+4\lambda\mathcal{A}\mathcal{S}>0$ we have that the motion is bounded, satisfying $|Q|\leq\sqrt{2\mathcal{H}/(\Omega^{2}+4\lambda\mathcal{A}\mathcal{S})}$ and $|\dot{Q}|\leq\sqrt{2\mathcal{H}}$. These bounds are obtained from  dropping either the kinetic or potential energy contributions in $\mathcal{H}$.

As a final example, we  consider a more sophisticated model described in Ref.\cite{SalaTufarelli2018} which includes in the radiation pressure hamiltonian corrections to the dependence on the wall's motion. This way, their model integrates all the features just mentioned for the previous cases.

\subsection{The case $f_{\alpha}\propto Q$ with photon creation and annihilation terms}\label{SectFQ}

At the next level of increasing complexity we consider an interaction function such that $f_{\alpha}\neq 0$. Unlike the previous case, it can be foreseen that the effective equation of motion will include dissipation and noise terms. The type of differential equation, what kind of dissipation and noise, will vary with the specific forms of  its functional dependence on $Q$. The simplest case corresponds to the linear case, where $f_{\alpha}=\lambda C_{\alpha}Q$, with $C_{\alpha}=C_{\alpha,v}-C_{\alpha,u}$ some constant related to the specific interaction hamiltonian. In terms of the interaction action of Eq.(\ref{SintFull}), we  set  $v_{\alpha}=\lambda C_{\alpha,v}Q$ and $u_{\alpha}=\lambda C_{\alpha,u}Q$, implying an asymmetry in the strength of the interactions with position and momentum of the cavity modes. Given this, the dissipation and noise terms read:
\begin{equation}
\int_{0}^{s}d\tau~\widetilde{\gamma}_{\alpha}(s,\tau)f'_{\alpha}\left(Q(s)\right)f'_{\alpha}\left(Q(\tau)\right)\dot{Q}(\tau)=\lambda^{2}\int_{0}^{s}d\tau~C_{\alpha}^{2}\widetilde{\gamma}_{\alpha}(s,\tau)\dot{Q}(\tau)~~,~~f'_{\alpha}\left(Q(s)\right)\xi_{\alpha}(s)=\lambda C_{\alpha}\xi_{\alpha}(s).
\end{equation}

Notice that while the dissipation term results in a typical linear dissipation term proportional to the velocity $\dot{Q}$, the noise is independent of $Q$, which corresponds to an additive noise.

Consider the model in Ref.\cite{Macri2018} for comparison. {As mentioned in Sect.\ref{SectQOMModels},} without discarding the dynamical Casimir effect terms accounting for the creation and annihilation of photon pairs, the interaction hamiltonian has the form $\hat{H}_{\rm Int}=\omega^{2}\hat{q}^{2}\lambda C\hat{Q}$, using the notation in Ref.\cite{ChoHu26}, where the coupling constant $C=mG/\omega$, having $G$ as the coupling parameter in Ref.\cite{Macri2018}. This model considers the first order in the interaction between the cavity mode and the wall given by the radiation pressure. This is particularly accurate for small displacements of the wall.

Considering the form of the interaction action of Eq.(\ref{SintFull}), we have that only one term is considered, matching with the interaction hamiltonian $H_{\rm Int}$ by setting $v_{nm}(Q)=\delta_{n,1}\delta_{nm}\lambda CQ$ and $u_{nm}(Q)=0$. This implies that there is no interaction with momentum, standing as a straightforward example of the asymmetry in the interactions mentioned before. In this sense, our modeling and analysis go further than what is required in the specific model,  since no interaction with the momentum is present.  In fact, this case can be addressed by the method developed in Ref.\cite{HPZ93}, where coupling terms involving just position operators were considered.

Here, only a single $f_{\alpha}$ is required, such that $f_{\alpha}(Q)=\lambda CQ$. Then, the corrections to the potential are $\delta V_{+}^{(1)}(Q)
=-\lambda\hbar\omega CQ/(2m)$ and $\delta V_{ab+}^{(2)}(Q)=0$. Accordingly, only one stochastic variable is introduced. Thus, the nonlinear Langevin equation reads:
\begin{equation}
M\ddot{Q}+M\Omega^{2}Q+\lambda^{2}C^{2}\int_{0}^{s}d\tau\widetilde{\gamma}(s,\tau)\dot{Q}(\tau)=\lambda C\xi(s)+\lambda\frac{\hbar\omega C}{2m}.
\label{QBrownian}
\end{equation}

Given the simplicity of the interaction model, the Langevin equation up to second order in the interaction which results is linear. The equation is formally the same as the one for a Brownian oscillator interacting with a single bath plus a constant applied force. In this sense, dissipation, noise and effective forcing are obtained in the approximated dynamics. Moreover, the convolution term accounting for the dissipation introduces memory effects in the dynamics of the system even in the linear case.

Before proceeding to the next section, it is worth noting here that a quadratic dependence of $f_{\alpha}$ would eventually give rise to multiplicative noises and nonlinear dissipation terms as in the dynamics of a van der Pol oscillator.

\subsection{The nonlinear case for $f_{\alpha}$ and models beyond wall's linear displacement}\label{SectFNonLinear}

Until this point we have analyzed optomechanical models based on radiation pressure and small amplitude motion of the wall.  However, several works go beyond this linear approximation and deal with optomechanical models where the motion of the wall are described by higher orders in the displacement amplitude.

{As previously mentioned in Sect.\ref{SectQOMModels},} a recent work in this direction is Ref.\cite{Ferreri2025}, where the authors expand the optomechanical interaction up to third order in powers of the dimensionless amplitude $\epsilon=\delta L/L\ll 1$ in a Hamiltonian of the form $\hat{H}=\hat{H}_{0}+\epsilon\hat{H}_{1}+\epsilon^{2}\hat{H}_{2}+\epsilon^{3}\hat{H}_{3}$.  This allows to capture resonant interactions, the contribution of counterrotating terms, cross-Kerr frequency shifts and Kerr-like interactions at once.  For the 1D limit case, the form of the interaction terms gives an action of the form of Eq.(\ref{SintFull}) such that:
\begin{equation}
v_{nm}(Q)=(-1)^{n+m+1}\sqrt{m_{n}m_{m}}~\epsilon\frac{Q}{Q_{\rm zp}}\left(1-\epsilon\frac{Q}{Q_{\rm zp}}-\left[\frac{(k_{n}^{2}+k_{m}^{2})L^{2}}{6}-1\right]\left[\epsilon\frac{Q}{Q_{\rm zp}}\right]^{2}\right),
\end{equation}
\begin{equation}
u_{nm}(Q)=(-1)^{n+m}\sqrt{m_{n}m_{m}}\frac{k_{n}k_{m}L^{2}}{3}\left(\epsilon\frac{Q}{Q_{\rm zp}}\right)^{3}.
\end{equation}

Notice that, first, their model corresponds to having $f_{\alpha}\neq 0$ in our modeling, including interactions with position (given by $v_{nm}$) and momentum (associated with $u_{nm}$). Second, there are terms with different powers of $\epsilon$, implying different powers of $Q$ here, resulting in nonlinear interaction functions $f_{\alpha}(Q)$. Third, the phenomenology presented in this model stems from triple interaction between two cavity modes and the wall's degree of freedom, having cross terms for the cavity modes. All these features are contained in our more general modeling and approach, as can be seen from Eq.(\ref{SintFull}).

From what has been learned from the analysis of our more general model, we highlight one useful insight, namely,  that each $\{n,m\}$ is matched into a stochastic variable with $\alpha\leftrightarrow\{n,m\}$. For the present case, the interaction functions read:
\begin{equation}
f_{\{n,m\}}(Q)=(-1)^{n+m+1}\lambda\sqrt{m_{n}m_{m}}~\epsilon\frac{Q}{Q_{\rm zp}}\left(1-\epsilon\frac{Q}{Q_{\rm zp}}-\left[\frac{(k_{n}-k_{m})^{2}L^{2}}{6}-1\right]\left[\epsilon\frac{Q}{Q_{\rm zp}}\right]^{2}\right).
\label{fnmFerreri}
\end{equation}
Notice that for the terms having $n=m$, the interaction functions simplify to:
\begin{equation}
f_{\{n,n\}}(Q)=-\lambda m_{n}\epsilon\frac{Q}{Q_{\rm zp}}\left(1-\epsilon\frac{Q}{Q_{\rm zp}}\right),
\label{fnnMacri}
\end{equation}
{giving a lower degree of nonlinearity than the terms coming from the cross terms between cavity modes in Eq.(\ref{fnmFerreri}).}

In any case, all the interaction functions anticipate nonlinear contributions in the Langevin equation of motion due to the damping and noise terms but also because of the potential renormalizations. While the renormalizations and the noise terms appear local in time, the damping terms involve time integrals of nonlinear powers of $Q$ and its time derivative, accounting for memory effects in the dynamics, as we mentioned before. Furthermore, each noise term has three distinct type contributions: an additive one (as in Brownian motion), a multiplicative one appearing in the frequency of the oscillator and a nonlinear one associated with $Q^{2}$. For a given contribution associated with a specific $\alpha$, the three noises are correlated since they are given by the same stochastic variable $\xi_{\alpha}$.

By keeping the third order terms in $\epsilon$, the second order in $\lambda$ renormalization to the potentials $\delta V_{nm+}^{(2)}$ is neglected, while the first order $\delta V_{n+}^{(1)}$ is fully retained, yielding terms in the equation of motion up to $Q^{2}$. Moreover, the product in the damping terms for every $\{n,m\}$ read:
\begin{equation}
f'_{\{n,m\}}\left(Q(s)\right)f'_{\{n,m\}}\left(Q(\tau)\right)\approx\lambda^{2}m_{n}m_{m}\frac{\epsilon^{2}}{Q_{\rm zp}^{2}}\left(1-2\epsilon\frac{\left[Q(s)+Q(\tau)\right]}{Q_{\rm zp}}\right),
\end{equation}
which in the end is linear in $Q$. Thus, keeping the third order in displacement in this model, the damping terms are:
\begin{equation}
\int_{0}^{s}d\tau\sum_{\alpha=1}^{\frac{N(N+1)}{2}}\widetilde{\gamma}_{\alpha}(s,\tau)f'_{\alpha}\left(Q(s)\right)f'_{\alpha}\left(Q(\tau)\right)\dot{Q}(\tau)=\int_{0}^{s}d\tau~\widetilde{\gamma}_{\rm M}(s,\tau)\frac{\epsilon^{2}}{Q_{\rm zp}^{2}}\left(1-2\epsilon\frac{\left[Q(s)+Q(\tau)\right]}{Q_{\rm zp}}\right)\dot{Q}(\tau),
\end{equation}
where:
\begin{equation}
\widetilde{\gamma}_{\rm M}(s,\tau)=\lambda^{2}\sum_{\alpha=1}^{\frac{N(N+1)}{2}}m_{n}m_{m}\widetilde{\gamma}_{\alpha}(s,\tau),
\end{equation}
in a similar fashion as it was done in Eq.(\ref{NonlinearChoHuEquation}), but in this case including also the cross  terms.

All in all, a similar analysis can be performed for other optomechanical models such as the ones of Refs.\cite{ButeraHi,Butera26} after a proper redefinition of the momenta as the ones carried out in Ref.\cite{ButeraHi}. The general features are similar to the ones discussed earlier for the model of Ref.\cite{Ferreri2025}. 

Apart from all this, it is crucial to notice that an expansion in powers of the interaction strength $\lambda$ implies an approximation in addition to the expansion in $\epsilon$ for small displacements. As mentioned in Refs.\cite{ButeraHi,Butera26}, for the case analyzed in this section, $\lambda$ accounts for the strength of the interaction only and not for an expansion in the displacement, which is already controlled by $\epsilon$. The notion of the interaction strength  arises because field modes with frequencies higher than the plasma frequency $\omega_{\rm pl}$ of the movable wall cannot effectively couple to the wall itself. This formally prevents the sums over modes from ultraviolet divergences. In that sense, the expansion parameter for this case should be linked to a comparison between the plasma frequency and the natural frequency of the wall, i.e. $\lambda\sim\omega_{\rm pl}/\Omega$.

\section{Conclusion}\label{SecConclu}

In this work, we approach the important and rapidly growing field of quantum optomechanics from general grounds by employing an underused yet powerful approach based on the Schwinger-Keldysh formalism with the Feynman-Vernon influence functional method, in accounting for the opto-mechanical backreaction effects of the quantum field environment on the system of moving cavity walls made of mirrors, membranes or material media.  Considering the plethora of models found in the literature, involving multiple aspects and phenomenology of set-ups involving cavity modes interacting with movable walls, we aim at reaching the parameter ranges or domains not yet or not easily reachable by conventional methods, such as the popular radiation pressure approach with interactions proportional to the cavity modes' photon numbers. 

\noindent{\bf Scope}

Our model is grounded in quantum optomechanics but it can also deal with processes in dynamical Casimir effect such as the creation and annihilation of photon pairs \cite{Mey08}.  More importantly, it provides a way to go beyond the small displacement approximation necessary for most calculations hitherto.  We can treat arbitrary displacement of the mirrors or membranes, thus reaching out to physical effects like cross-Kerr frequency shifts and Kerr-like interactions, as in Ref.\cite{Ferreri2025}.   
This general model is capable of including at once multiple possible aspects and features of optomechanical systems, showing up in the compact form of the interaction action of Eq.(\ref{SintFull}).  Given the general form in terms of the interaction functions $v_{\alpha}$ and $u_{\alpha}$,  our model not only includes couplings between the wall's degree of freedom and specific cavity modes, but also triple couplings where two cavity modes are involved in the interaction with the wall.  The couplings involve quadratic products of either position-only or momentum-only cavity operators. Models with interaction terms involving products between position and momentum can be treated with suitable canonical transformations leading to position-only and momentum-only quadratic products, such as it is done, for instance, in Refs.\cite{ButeraHi,Butera26}.

\noindent{\bf Features}

Using the functional perturbative method of \cite{HPZ93,ChoHu26} under the assumption of weak opto-mechanical coupling or small $\lambda$,  we have obtained the effective action $S_{\rm IF}$ up to second order in $\lambda$ for the wall's degree of freedom. From this, we obtained the effective equation of motion in the form of a nonlinear Langevin equation Eq.(\ref{NonlinearLangevinEquation}).
An important observation about the perturbation parameter $\lambda$ in our model is that it encompasses both the case of small amplitudes of the wall's motion ($\lambda\sim\epsilon=\delta L/L$) or the weak interaction between the cavity modes and the wall ($\lambda\sim\omega_{\rm Pl}/\Omega$), as mentioned in Refs.\cite{ButeraHi,Butera26}. However, as we proceed to apply our results to Ref.\cite{Ferreri2025}, by keeping the perturbation parameter $\lambda$ associated with $\omega_{\rm Pl}/\Omega$, we have shown that our model is applicable beyond the approximation of small displacements, provided that the correct form of the interaction functions $v_{\alpha}$ and $u_{\alpha}$ for such regime are used. This is a striking insight since we are showing that the effective equation of motion for the wall's displacement up to second order in the strength of the interaction between the cavity modes and the wall has always the form of Eq.(\ref{NonlinearLangevinEquation}), regardless of the amplitude of the motion. In an analog manner, one might consider the situation where the perturbation parameter $\lambda$ is associated only with the displacement $\epsilon$, while the interaction functions $v_{\alpha}$ and $u_{\alpha}$ encode the information up to arbitrary orders in $\omega_{\rm Pl}/\Omega$. Also in this situation, Eq.(\ref{NonlinearLangevinEquation}) remains applicable. In this sense,  Eq.(\ref{NonlinearLangevinEquation}) is both fundamental and robust.

\noindent{\bf Dynamics}

To connect our general model with the many representative models in the literature where the forms and types of the effective equations of motion vary, we have analyzed the different possibilities by grouping them according to general features characterizing the models, namely, the case $f_{\alpha}=v_{\alpha}-u_{\alpha}=0$, the linear case $f_{\alpha}\propto Q$ and the nonlinear case for $f_{\alpha}$. We showed that each case presents a different phenomenology. For instance, the case $f_{\alpha}=0$ given in Sect.\ref{SectF0} is associated with interaction terms proportional to the number operator of a cavity mode $\hat{n}_{\alpha}$. In this case, linear and nonlinear equations can be obtained, but the striking feature is that the equations are local in time. For this case, the dynamics are Markovian, so no memory effects are evidenced in the equation, and no dissipation nor noise takes place. Nonlinear effects come from the renormalizations of the potential $\delta V_{+}$. In contrast, the case $f_{\alpha}\propto Q$ considered in Sect.\ref{SectFQ}, typically happening when creation and annihilation of photon pairs are considered, leads to a linear Langevin equation as in Brownian motion. In this case, dissipation and (additive) noise take place, and the dynamics can be non-Markovian in principle, depending on the damping kernels $\gamma_{\alpha}$ associated with each interaction term. Finally, the nonlinear case for $f_{\alpha}$ analyzed in Sect.\ref{SectFNonLinear} allows for obtaining multiplicative and more exotic noises entering the equation, balanced by the corresponding nonlinear damping terms.

\noindent{\bf Noise}

Another noteworthy gain in taking this approach and modeling is the ability to establish a correspondence between each interaction term with the stochastic variable accounting for independent noises. These noises could enter the dynamics in different ways, being additive, multiplicative or even of more exotic nature. Our framework formally treats interaction terms with a specific cavity mode and cross terms where two cavity modes are involved on equal footing. In the present treatment where the perturbative expansion is up to the second order in $\lambda$, the associated noises are Gaussian. At the third order expansion in $\lambda$, the noises will be non-Gaussian, as expounded in Ref.\cite{ChoHu26}.  At higher order expansions in $\lambda$,  new Langevin equations can be derived with some effort,  but it is worthwhile. Because, with careful analysis in making the connection to current experiments and measurement platforms, characterized by general interaction functions $v_{\alpha}$ and $u_{\alpha}$, these results can reach a broader range of applicability in QOM.

\begin{acknowledgments}
A.E.R.L. is supported by ANID through grants FONDECYT Iniciaci\'on No. 11250638 and the Millennium Science Initiative Program ICN17\_012. HTC was supported in part by the National Science and Technology Council (NSTC) of Taiwan, Republic of China, under Grant Nos.~NSTC 114-2112-M-032-007 and NSTC 115-2112-M-032-005.
\end{acknowledgments}

\bibliography{Biblio}

\appendix

\section{Functional derivatives of the environmental generating functional}\label{AppFunctionalDerivatives}

This Appendix is devoted to the calculation of the functional derivatives required for the influence action up to second order in the opto-mechanical coupling  $\lambda$.

We first write the functional derivative of the generating functional $Z$ as:
\begin{equation}
\frac{\delta Z}{\delta J_{b\pm}(\tau)}=Y_{b\pm}(\tau)Z,
\end{equation}
\begin{eqnarray}
Y_{b\pm}(\tau)&=&\frac{i}{2\hbar^{2}}\int_{0}^{t}dt'\left[G_{b\pm\pm}(\tau,t')J_{b\pm}(t')+J_{b\pm}(t')G_{b\pm\pm}(t',\tau)-G_{b\pm\mp}(\tau,t')J_{b\mp}(t')\right.\\
&&\left.-J_{b\mp}(t')G_{b\mp\pm}(t',\tau)\right],\nonumber
\end{eqnarray}
\begin{equation}
\frac{\delta Y_{b\pm}(\tau)}{\delta J_{a\pm}(\tau')}=\frac{i}{\hbar^{2}}\delta_{ab}G_{b\pm\pm}(\tau,\tau')~~,~~\frac{\delta Y_{b\pm}(\tau)}{\delta J_{a\mp}(\tau')}=-\frac{i}{\hbar^{2}}\delta_{ab}G_{b\pm\mp}(\tau,\tau'),
\end{equation}

Given these expressions, we have that:
\begin{equation}
\frac{\delta}{\delta J_{a\pm}(\tau')}\frac{\delta Z}{\delta J_{b\pm}(\tau)}=\frac{i}{\hbar^{2}}\delta_{ab}G_{b\pm\pm}(\tau,\tau')Z+Y_{a\pm}(\tau')Y_{b\pm}(\tau)Z.
\end{equation}

At this point, we divide terms into two groups, we proceed first with the derivatives of the same subscript sign:
\begin{eqnarray}
\frac{\delta}{\delta J_{m\pm}(\tau'')}\frac{\delta}{\delta J_{a\pm}(\tau')}\frac{\delta Z}{\delta J_{b\pm}(\tau)}&=&\frac{i}{\hbar^{2}}\delta_{ab}G_{b\pm\pm}(\tau,\tau')Y_{m\pm}(\tau'')Z+\frac{i}{\hbar^{2}}\delta_{ma}G_{a\pm\pm}(\tau',\tau'')Y_{b\pm}(\tau)Z\nonumber\\
&+&\frac{i}{\hbar^{2}}\delta_{mb}G_{b\pm\pm}(\tau,\tau'')Y_{a\pm}(\tau')Z+Y_{m\pm}(\tau'')Y_{a\pm}(\tau')Y_{b\pm}(\tau)Z,
\end{eqnarray}
which  gives:
\begin{eqnarray}
&&\frac{\delta}{\delta J_{n\pm}(\tau''')}\frac{\delta}{\delta J_{m\pm}(\tau'')}\frac{\delta}{\delta J_{a\pm}(\tau')}\frac{\delta Z}{\delta J_{b\pm}(\tau)}=\\
&&=\frac{i}{\hbar^{2}}\delta_{ab}G_{b\pm\pm}(\tau,\tau')\frac{i}{\hbar^{2}}\delta_{nm}G_{m\pm\pm}(\tau'',\tau''')Z+\frac{i}{\hbar^{2}}\delta_{ab}G_{b\pm\pm}(\tau,\tau')Y_{m\pm}(\tau'')Y_{n\pm}(\tau''')Z\nonumber\\
&&+\frac{i}{\hbar^{2}}\delta_{ma}G_{a\pm\pm}(\tau',\tau'')\frac{i}{\hbar^{2}}\delta_{nb}G_{b\pm\pm}(\tau,\tau''')Z+\frac{i}{\hbar^{2}}\delta_{ma}G_{a\pm\pm}(\tau',\tau'')Y_{b\pm}(\tau)Y_{n\pm}(\tau''')Z\nonumber\\
&&+\frac{i}{\hbar^{2}}\delta_{mb}G_{b\pm\pm}(\tau,\tau'')\frac{i}{\hbar^{2}}\delta_{na}G_{a\pm\pm}(\tau',\tau''')Z+\frac{i}{\hbar^{2}}\delta_{mb}G_{b\pm\pm}(\tau,\tau'')Y_{a\pm}(\tau')Y_{n\pm}(\tau''')Z\nonumber\\
&&+\frac{i}{\hbar^{2}}\delta_{nm}G_{m\pm\pm}(\tau'',\tau''')Y_{a\pm}(\tau')Y_{b\pm}(\tau)Z+Y_{m\pm}(\tau'')\frac{i}{\hbar^{2}}\delta_{na}G_{a\pm\pm}(\tau',\tau''')Y_{b\pm}(\tau)Z\nonumber\\
&&+Y_{m\pm}(\tau'')Y_{a\pm}(\tau')\frac{i}{\hbar^{2}}\delta_{nb}G_{b\pm\pm}(\tau,\tau''')Z+Y_{n\pm}(\tau''')Y_{m\pm}(\tau'')Y_{a\pm}(\tau')Y_{b\pm}(\tau)Z,
\end{eqnarray}
such that:
\begin{eqnarray}
&&\left.\frac{\delta}{\delta J_{n\pm}(\tau''')}\frac{\delta}{\delta J_{m\pm}(\tau'')}\frac{\delta}{\delta J_{a\pm}(\tau')}\frac{\delta Z}{\delta J_{b\pm}(\tau)}\right|_{J_{n+}=J_{n-}=0}=\\
&&=-\frac{1}{\hbar^{4}}\left[\delta_{ab}G_{b\pm\pm}(\tau,\tau')\delta_{nm}G_{m\pm\pm}(\tau'',\tau''')+\delta_{ma}G_{a\pm\pm}(\tau',\tau'')\delta_{nb}G_{b\pm\pm}(\tau,\tau''')\right.\nonumber\\
&&\left.+\delta_{mb}G_{b\pm\pm}(\tau,\tau'')\delta_{na}G_{a\pm\pm}(\tau',\tau''')\right].\nonumber
\end{eqnarray}

From this, we can obtain the following derivatives that are required for our calculation:
\begin{eqnarray}
&&\left.\frac{\delta}{\delta J_{n\pm}(s)}\frac{\delta}{\delta J_{m\pm}(s)}\frac{\delta}{\delta J_{a\pm}(s')}\frac{\delta Z}{\delta J_{b\pm}(s')}\right|_{J_{n+}=J_{n-}=0}=\\
&&=-\frac{1}{\hbar^{4}}\left[\delta_{ab}\delta_{nm}G_{b\pm\pm}(s',s')G_{m\pm\pm}(s,s)+\left(\delta_{ma}\delta_{nb}+\delta_{mb}\delta_{na}\right)G_{a\pm\pm}(s',s)G_{b\pm\pm}(s',s)\right].\nonumber
\end{eqnarray}

For the other group, we have:
\begin{eqnarray}
\frac{\delta}{\delta J_{m\mp}(\tau'')}\frac{\delta}{\delta J_{a\pm}(\tau')}\frac{\delta Z}{\delta J_{b\pm}(\tau)}&=&\frac{i}{\hbar^{2}}\delta_{ab}G_{b\pm\pm}(\tau,\tau')Y_{m\mp}(\tau'')Z-\frac{i}{\hbar^{2}}\delta_{ma}G_{a\pm\mp}(\tau',\tau'')Y_{b\pm}(\tau)Z\nonumber\\
&-&\frac{i}{\hbar^{2}}\delta_{mb}G_{b\pm\mp}(\tau,\tau'')Y_{a\pm}(\tau')Z+Y_{m\mp}(\tau'')Y_{a\pm}(\tau')Y_{b\pm}(\tau)Z,
\end{eqnarray}
which then gives:
\begin{eqnarray}
&&\frac{\delta}{\delta J_{n\mp}(\tau''')}\frac{\delta}{\delta J_{m\mp}(\tau'')}\frac{\delta}{\delta J_{a\pm}(\tau')}\frac{\delta Z}{\delta J_{b\pm}(\tau)}=\nonumber\\
&&=\frac{i}{\hbar^{2}}\delta_{ab}G_{b\pm\pm}(\tau,\tau')\frac{i}{\hbar^{2}}\delta_{nm}G_{m\mp\mp}(\tau'',\tau''')Z+\frac{i}{\hbar^{2}}\delta_{ab}G_{b\pm\pm}(\tau,\tau')Y_{m\pm}(\tau'')Y_{n\mp}(\tau''')Z\nonumber\\
&&+\frac{i}{\hbar^{2}}\delta_{ma}G_{a\pm\mp}(\tau',\tau'')\frac{i}{\hbar^{2}}\delta_{nb}G_{b\pm\mp}(\tau,\tau''')Z-\frac{i}{\hbar^{2}}\delta_{ma}G_{a\pm\mp}(\tau',\tau'')Y_{b\pm}(\tau)Y_{n\mp}(\tau''')Z\nonumber\\
&&+\frac{i}{\hbar^{2}}\delta_{mb}G_{b\pm\mp}(\tau,\tau'')\frac{i}{\hbar^{2}}\delta_{na}G_{a\pm\mp}(\tau',\tau''')Z-\frac{i}{\hbar^{2}}\delta_{mb}G_{b\pm\mp}(\tau,\tau'')Y_{a\pm}(\tau')Y_{n\mp}(\tau''')Z\nonumber\\
&&+\frac{i}{\hbar^{2}}\delta_{nm}G_{m\mp\mp}(\tau'',\tau''')Y_{a\pm}(\tau')Y_{b\pm}(\tau)Z-Y_{m\mp}(\tau'')\frac{i}{\hbar^{2}}\delta_{na}G_{a\pm\mp}(\tau',\tau''')Y_{b\pm}(\tau)Z\nonumber\\
&&-Y_{m\mp}(\tau'')Y_{a\pm}(\tau')\frac{i}{\hbar^{2}}\delta_{nb}G_{b\pm\mp}(\tau,\tau''')Z+Y_{n\mp}(\tau''')Y_{m\mp}(\tau'')Y_{a\pm}(\tau')Y_{b\pm}(\tau)Z,
\end{eqnarray}
such that:
\begin{eqnarray}
&&\left.\frac{\delta}{\delta J_{n\mp}(\tau''')}\frac{\delta}{\delta J_{m\mp}(\tau'')}\frac{\delta}{\delta J_{a\pm}(\tau')}\frac{\delta Z}{\delta J_{b\pm}(\tau)}\right|_{J_{n+}=J_{n-}=0}=\\
&&=-\frac{1}{\hbar^{4}}\left[\delta_{ab}G_{b\pm\pm}(\tau,\tau')\delta_{nm}G_{m\mp\mp}(\tau'',\tau''')+\delta_{ma}G_{a\pm\mp}(\tau',\tau'')\delta_{nb}G_{b\pm\mp}(\tau,\tau''')\right.\nonumber\\
&&\left.+\delta_{mb}G_{b\pm\mp}(\tau,\tau'')\delta_{na}G_{a\pm\mp}(\tau',\tau''')\right].\nonumber
\end{eqnarray}

From this, we get:
\begin{eqnarray}
&&\left.\frac{\delta}{\delta J_{n\mp}(s)}\frac{\delta}{\delta J_{m\mp}(s)}\frac{\delta}{\delta J_{a\pm}(s')}\frac{\delta Z}{\delta J_{b\pm}(s')}\right|_{J_{n+}=J_{n-}=0}=\\
&&=-\frac{1}{\hbar^{4}}\left[\delta_{ab}\delta_{nm}G_{b\pm\pm}(s',s')G_{m\mp\mp}(s,s)+\left(\delta_{ma}\delta_{nb}+\delta_{mb}\delta_{na}\right)G_{a\pm\mp}(s',s)G_{b\pm\mp}(s',s)\right].\nonumber
\end{eqnarray}

All these expressions are required for the calculation of the influence action up to the second order in $\lambda$.

\section{Calculation of the second order influence action}\label{App2ndIF}

To calculate the second order  influence action we start from Eq.(\ref{SIF2}) with this consideration:
\begin{eqnarray}
&&S_{\rm int}\left[Q_{+},\left\{\frac{\hbar}{i}\frac{\delta}{\delta J_{n+}}\right\}\right]-S_{\rm int}\left[Q_{-},\left\{-\frac{\hbar}{i}\frac{\delta}{\delta J_{n-}}\right\}\right]=\\
&&=-\hbar^{2}\int_{0}^{t}ds\sum_{n,m}\left.\left(v_{nm}(Q_{+})\omega_{n}\omega_{m}~\frac{\delta}{\delta J_{n+}(s)}\frac{\delta}{\delta J_{m+}(s)}+u_{nm}(Q_{+})\left.\partial^{2}_{\tau\tau'}\frac{\delta}{\delta J_{n+}(\tau)}\frac{\delta}{\delta J_{m+}(\tau')}\right|_{\tau=\tau'=s}\right.\right.\nonumber\\
&&-\left.v_{nm}(Q_{-})\omega_{n}\omega_{m}~\frac{\delta}{\delta J_{n-}(s)}\frac{\delta}{\delta J_{m-}(s)}-u_{nm}(Q_{-})\left.\partial^{2}_{\tau\tau'}\frac{\delta}{\delta J_{n-}(\tau)}\frac{\delta}{\delta J_{m-}(\tau')}\right|_{\tau=\tau'=s}\right),\nonumber
\end{eqnarray}
in such a way that:
\begin{eqnarray}
&&\left(S_{\rm int}\left[Q_{+},\left\{\frac{\hbar}{i}\frac{\delta}{\delta J_{n+}}\right\}\right]-S_{\rm int}\left[Q_{-},\left\{-\frac{\hbar}{i}\frac{\delta}{\delta J_{n-}}\right\}\right]\right)^{2}=\hbar^{4}\int_{0}^{t}ds\int_{0}^{t}ds'\sum_{n,m,a,b}\\
&&\times\left(v_{nm}(Q_{+})\omega_{n}\omega_{m}~\frac{\delta}{\delta J_{n+}(s)}\frac{\delta}{\delta J_{m+}(s)}+u_{nm}(Q_{+})\left.\partial^{2}_{\tau'''\tau''}\frac{\delta}{\delta J_{n+}(\tau''')}\frac{\delta}{\delta J_{m+}(\tau'')}\right|_{\tau''=\tau'''=s}\right.\nonumber\\
&&-\left.v_{nm}(Q_{-})\omega_{n}\omega_{m}~\frac{\delta}{\delta J_{n-}(s)}\frac{\delta}{\delta J_{m-}(s)}-u_{nm}(Q_{-})\left.\partial^{2}_{\tau'''\tau''}\frac{\delta}{\delta J_{n-}(\tau''')}\frac{\delta}{\delta J_{m-}(\tau'')}\right|_{\tau''=\tau'''=s}\right)\nonumber\\
&&\times\left.\left(v_{ab}(Q_{+})\omega_{a}\omega_{b}~\frac{\delta}{\delta J_{a+}(s')}\frac{\delta}{\delta J_{b+}(s')}+u_{ab}(Q_{+})\left.\partial^{2}_{\tau'\tau}\frac{\delta}{\delta J_{a+}(\tau')}\frac{\delta}{\delta J_{b+}(\tau)}\right|_{\tau=\tau'=s'}\right.\right.\nonumber\\
&&-\left.v_{ab}(Q_{-})\omega_{a}\omega_{b}~\frac{\delta}{\delta J_{a-}(s')}\frac{\delta}{\delta J_{b-}(s')}-u_{ab}(Q_{-})\left.\partial^{2}_{\tau'\tau}\frac{\delta}{\delta J_{a-}(\tau')}\frac{\delta}{\delta J_{b-}(\tau)}\right|_{\tau=\tau'=s'}\right),\nonumber
\end{eqnarray}
with the notation  $u,v_{nm}$ appearing  on the left (right) are evaluated at $s$ ($s'$). Gathering all the products explicitly:
\begin{eqnarray}
&&\left(S_{\rm int}\left[Q_{+},\left\{\frac{\hbar}{i}\frac{\delta}{\delta J_{n+}}\right\}\right]-S_{\rm int}\left[Q_{-},\left\{-\frac{\hbar}{i}\frac{\delta}{\delta J_{n-}}\right\}\right]\right)^{2}=\\
&&=\hbar^{4}\int_{0}^{t}ds\int_{0}^{t}ds'\sum_{n,m,a,b}\left(v_{nm}(Q_{+})\omega_{n}\omega_{m}v_{ab}(Q_{+})\omega_{a}\omega_{b}~\frac{\delta}{\delta J_{n+}(s)}\frac{\delta}{\delta J_{m+}(s)}\frac{\delta}{\delta J_{a+}(s')}\frac{\delta}{\delta J_{b+}(s')}\right.\nonumber\\
&&+\left.v_{nm}(Q_{+})\omega_{n}\omega_{m}u_{ab}(Q_{+})~\frac{\delta}{\delta J_{n+}(s)}\frac{\delta}{\delta J_{m+}(s)}\left.\partial^{2}_{\tau'\tau}\frac{\delta}{\delta J_{a+}(\tau')}\frac{\delta}{\delta J_{b+}(\tau)}\right|_{\tau=\tau'=s'}\right.\nonumber\\
&&-\left.v_{nm}(Q_{+})\omega_{n}\omega_{m}v_{ab}(Q_{-})\omega_{a}\omega_{b}~\frac{\delta}{\delta J_{n+}(s)}\frac{\delta}{\delta J_{m+}(s)}\frac{\delta}{\delta J_{a-}(s')}\frac{\delta}{\delta J_{b-}(s')}\right.\nonumber\\
&&-\left.v_{nm}(Q_{+})\omega_{n}\omega_{m}u_{ab}(Q_{-})~\frac{\delta}{\delta J_{n+}(s)}\frac{\delta}{\delta J_{m+}(s)}\left.\partial^{2}_{\tau'\tau}\frac{\delta}{\delta J_{a-}(\tau')}\frac{\delta}{\delta J_{b-}(\tau)}\right|_{\tau=\tau'=s'}\right.\nonumber\\
&&+\left.\left.u_{nm}(Q_{+})v_{ab}(Q_{+})\omega_{a}\omega_{b}\left.\partial^{2}_{\tau'''\tau''}\frac{\delta}{\delta J_{n+}(\tau''')}\frac{\delta}{\delta J_{m+}(\tau'')}\right|_{\tau''=\tau'''=s}\frac{\delta}{\delta J_{a+}(s')}\frac{\delta}{\delta J_{b+}(s')}\right.\right.\nonumber\\
&&+\left.\left.u_{nm}(Q_{+})u_{ab}(Q_{+})\left.\partial^{2}_{\tau'''\tau''}\frac{\delta}{\delta J_{n+}(\tau''')}\frac{\delta}{\delta J_{m+}(\tau'')}\right|_{\tau''=\tau'''=s}\left.\partial^{2}_{\tau'\tau}\frac{\delta}{\delta J_{a+}(\tau')}\frac{\delta}{\delta J_{b+}(\tau)}\right|_{\tau=\tau'=s'}\right.\right.\nonumber\\
&&-\left.\left.u_{nm}(Q_{+})v_{ab}(Q_{-})\omega_{a}\omega_{b}\left.\partial^{2}_{\tau'''\tau''}\frac{\delta}{\delta J_{n+}(\tau''')}\frac{\delta}{\delta J_{m+}(\tau'')}\right|_{\tau''=\tau'''=s}\frac{\delta}{\delta J_{a-}(s')}\frac{\delta}{\delta J_{b-}(s')}\right.\right.\nonumber\\
&&-\left.\left.u_{nm}(Q_{+})u_{ab}(Q_{-})\left.\partial^{2}_{\tau'''\tau''}\frac{\delta}{\delta J_{n+}(\tau''')}\frac{\delta}{\delta J_{m+}(\tau'')}\right|_{\tau''=\tau'''=s}\left.\partial^{2}_{\tau'\tau}\frac{\delta}{\delta J_{a-}(\tau')}\frac{\delta}{\delta J_{b-}(\tau)}\right|_{\tau=\tau'=s'}\right.\right.\nonumber\\
&&-\left.v_{nm}(Q_{-})\omega_{n}\omega_{m}v_{ab}(Q_{+})\omega_{a}\omega_{b}~\frac{\delta}{\delta J_{n-}(s)}\frac{\delta}{\delta J_{m-}(s)}\frac{\delta}{\delta J_{a+}(s')}\frac{\delta}{\delta J_{b+}(s')}\right.\nonumber\\
&&-\left.v_{nm}(Q_{-})\omega_{n}\omega_{m}u_{ab}(Q_{+})~\frac{\delta}{\delta J_{n-}(s)}\frac{\delta}{\delta J_{m-}(s)}\left.\partial^{2}_{\tau'\tau}\frac{\delta}{\delta J_{a+}(\tau')}\frac{\delta}{\delta J_{b+}(\tau)}\right|_{\tau=\tau'=s'}\right.\nonumber\\
&&+\left.v_{nm}(Q_{-})\omega_{n}\omega_{m}v_{ab}(Q_{-})\omega_{a}\omega_{b}~\frac{\delta}{\delta J_{n-}(s)}\frac{\delta}{\delta J_{m-}(s)}\frac{\delta}{\delta J_{a-}(s')}\frac{\delta}{\delta J_{b-}(s')}\right.\nonumber\\
&&+\left.v_{nm}(Q_{-})\omega_{n}\omega_{m}u_{ab}(Q_{-})~\frac{\delta}{\delta J_{n-}(s)}\frac{\delta}{\delta J_{m-}(s)}\left.\partial^{2}_{\tau'\tau}\frac{\delta}{\delta J_{a-}(\tau')}\frac{\delta}{\delta J_{b-}(\tau)}\right|_{\tau=\tau'=s'}\right.\nonumber\\
&&-\left.u_{nm}(Q_{-})v_{ab}(Q_{+})\omega_{a}\omega_{b}\left.\partial^{2}_{\tau'''\tau''}\frac{\delta}{\delta J_{n-}(\tau''')}\frac{\delta}{\delta J_{m-}(\tau'')}\right|_{\tau''=\tau'''=s}\frac{\delta}{\delta J_{a+}(s')}\frac{\delta}{\delta J_{b+}(s')}\right.\nonumber\\
&&-\left.u_{nm}(Q_{-})u_{ab}(Q_{+})\left.\partial^{2}_{\tau'''\tau''}\frac{\delta}{\delta J_{n-}(\tau''')}\frac{\delta}{\delta J_{m-}(\tau'')}\right|_{\tau''=\tau'''=s}\left.\partial^{2}_{\tau'\tau}\frac{\delta}{\delta J_{a+}(\tau')}\frac{\delta}{\delta J_{b+}(\tau)}\right|_{\tau=\tau'=s'}\right.\nonumber\\
&&+\left.u_{nm}(Q_{-})v_{ab}(Q_{-})\omega_{a}\omega_{b}\left.\partial^{2}_{\tau'''\tau''}\frac{\delta}{\delta J_{n-}(\tau''')}\frac{\delta}{\delta J_{m-}(\tau'')}\right|_{\tau''=\tau'''=s}\frac{\delta}{\delta J_{a-}(s')}\frac{\delta}{\delta J_{b-}(s')}\right.\nonumber\\
&&+\left.u_{nm}(Q_{-})u_{ab}(Q_{-})\left.\partial^{2}_{\tau'''\tau''}\frac{\delta}{\delta J_{n-}(\tau''')}\frac{\delta}{\delta J_{m-}(\tau'')}\right|_{\tau''=\tau'''=s}\left.\partial^{2}_{\tau'\tau}\frac{\delta}{\delta J_{a-}(\tau')}\frac{\delta}{\delta J_{b-}(\tau)}\right|_{\tau=\tau'=s'}\right).\nonumber
\end{eqnarray}

Applying the expressions for the derivatives in the previous Appendix and considering Eq.(\ref{SIF1G}), we see that the required combination for the second order in Eq.(\ref{SIFExpansion}) corresponds to the substraction of the terms associated with $\delta_{ab}\delta_{nm}$ in the last expression. This is equivalent to just considering the connected terms, as done in Ref.\cite{ChoHu26}. Then, we have:
{
\fontsize{9.5pt}{12pt}\selectfont
\begin{eqnarray}
&&S_{\rm IF}^{(2)}\left[Q_{+},Q_{-}\right]=-\frac{i}{2\hbar}\int_{0}^{t}ds\int_{0}^{t}ds'\sum_{n,m,a,b}\left(\delta_{ma}\delta_{nb}+\delta_{mb}\delta_{na}\right)\\
&&\left[v_{nm}(Q_{+})\omega_{n}\omega_{m}v_{ab}(Q_{+})\omega_{a}\omega_{b}G_{a++}(s',s)G_{b++}(s',s)+v_{nm}(Q_{+})\omega_{n}\omega_{m}u_{ab}(Q_{+})\partial_{1}G_{a++}(s',s)\partial_{1}G_{b++}(s',s)\right.\nonumber\\
&&-\left.v_{nm}(Q_{+})\omega_{n}\omega_{m}v_{ab}(Q_{-})\omega_{a}\omega_{b}G_{a-+}(s',s)G_{b-+}(s',s)-v_{nm}(Q_{+})\omega_{n}\omega_{m}u_{ab}(Q_{-})\partial_{1}G_{a-+}(s',s)\partial_{1}G_{b-+}(s',s)\right.\nonumber\\
&&+\left.u_{nm}(Q_{+})v_{ab}(Q_{+})\omega_{a}\omega_{b}\partial_{2}G_{a++}(s',s)\partial_{2}G_{b++}(s',s)+u_{nm}(Q_{+})u_{ab}(Q_{+})\partial_{12}^{2}G_{a++}(s',s)\partial_{12}^{2}G_{b++}(s',s)\right.\nonumber\\
&&-\left.u_{nm}(Q_{+})v_{ab}(Q_{-})\omega_{a}\omega_{b}\partial_{2}G_{a-+}(s',s)\partial_{2}G_{b-+}(s',s)-u_{nm}(Q_{+})u_{ab}(Q_{-})\partial_{12}^{2}G_{a-+}(s',s)\partial_{12}^{2}G_{b-+}(s',s)\right.\nonumber\\
&&-\left.v_{nm}(Q_{-})\omega_{n}\omega_{m}v_{ab}(Q_{+})\omega_{a}\omega_{b}G_{a+-}(s',s)G_{b+-}(s',s)-v_{nm}(Q_{-})\omega_{n}\omega_{m}u_{ab}(Q_{+})\partial_{1}G_{a+-}(s',s)\partial_{1}G_{b+-}(s',s)\right.\nonumber\\
&&+\left.v_{nm}(Q_{-})\omega_{n}\omega_{m}v_{ab}(Q_{-})\omega_{a}\omega_{b}G_{a--}(s',s)G_{b--}(s',s)+v_{nm}(Q_{-})\omega_{n}\omega_{m}u_{ab}(Q_{-})\partial_{1}G_{a--}(s',s)\partial_{1}G_{b--}(s',s)\right.\nonumber\\
&&-\left.u_{nm}(Q_{-})v_{ab}(Q_{+})\omega_{a}\omega_{b}\partial_{2}G_{a+-}(s',s)\partial_{2}G_{b+-}(s',s)-u_{nm}(Q_{-})u_{ab}(Q_{+})\partial_{12}^{2}G_{a+-}(s',s)\partial_{12}^{2}G_{b+-}(s',s)\right.\nonumber\\
&&+\left.u_{nm}(Q_{-})v_{ab}(Q_{-})\omega_{a}\omega_{b}\partial_{2}G_{a--}(s',s)\partial_{2}G_{b--}(s',s)+u_{nm}(Q_{-})u_{ab}(Q_{-})\partial_{12}^{2}G_{a--}(s',s)\partial_{12}^{2}G_{b--}(s',s)\right].\nonumber
\end{eqnarray}
}

Considering that $u_{nm}=u_{mn}$ and $v_{nm}=v_{mn}$, while all the terms have the form $f_{mn}A_{abnm}g_{ab}$, where $f,g=u,v$, and $A_{abnm}=A_{banm}$, we have that the sums over $n,m,a,b$ including the factor $\delta_{ma}\delta_{nb}$ is the same as the one including $\delta_{mb}\delta_{na}$. Thus, we can write:
{
\fontsize{10pt}{12pt}\selectfont
\begin{eqnarray}
&&S_{\rm IF}^{(2)}\left[Q_{+},Q_{-}\right]=-\frac{i}{\hbar}\int_{0}^{t}ds\int_{0}^{t}ds'\sum_{a,b}\\
&&\left[v_{ba}(Q_{+})v_{ab}(Q_{+})\omega_{a}^{2}\omega_{b}^{2}G_{a++}(s',s)G_{b++}(s',s)+v_{ba}(Q_{+})\omega_{b}\omega_{a}u_{ab}(Q_{+})\partial_{1}G_{a++}(s',s)\partial_{1}G_{b++}(s',s)\right.\nonumber\\
&&-\left.v_{ba}(Q_{+})v_{ab}(Q_{-})\omega_{a}^{2}\omega_{b}^{2}G_{a-+}(s',s)G_{b-+}(s',s)-v_{ba}(Q_{+})\omega_{b}\omega_{a}u_{ab}(Q_{-})\partial_{1}G_{a-+}(s',s)\partial_{1}G_{b-+}(s',s)\right.\nonumber\\
&&+\left.u_{ba}(Q_{+})v_{ab}(Q_{+})\omega_{a}\omega_{b}\partial_{2}G_{a++}(s',s)\partial_{2}G_{b++}(s',s)+u_{ba}(Q_{+})u_{ab}(Q_{+})\partial_{12}^{2}G_{a++}(s',s)\partial_{12}^{2}G_{b++}(s',s)\right.\nonumber\\
&&-\left.u_{ba}(Q_{+})v_{ab}(Q_{-})\omega_{a}\omega_{b}\partial_{2}G_{a-+}(s',s)\partial_{2}G_{b-+}(s',s)-u_{ba}(Q_{+})u_{ab}(Q_{-})\partial_{12}^{2}G_{a-+}(s',s)\partial_{12}^{2}G_{b-+}(s',s)\right.\nonumber\\
&&-\left.v_{ba}(Q_{-})v_{ab}(Q_{+})\omega_{a}^{2}\omega_{b}^{2}G_{a+-}(s',s)G_{b+-}(s',s)-v_{ba}(Q_{-})\omega_{b}\omega_{a}u_{ab}(Q_{+})\partial_{1}G_{a+-}(s',s)\partial_{1}G_{b+-}(s',s)\right.\nonumber\\
&&+\left.v_{ba}(Q_{-})v_{ab}(Q_{-})\omega_{a}^{2}\omega_{b}^{2}G_{a--}(s',s)G_{b--}(s',s)+v_{ba}(Q_{-})\omega_{b}\omega_{a}u_{ab}(Q_{-})\partial_{1}G_{a--}(s',s)\partial_{1}G_{b--}(s',s)\right.\nonumber\\
&&-\left.u_{ba}(Q_{-})v_{ab}(Q_{+})\omega_{a}\omega_{b}\partial_{2}G_{a+-}(s',s)\partial_{2}G_{b+-}(s',s)-u_{ba}(Q_{-})u_{ab}(Q_{+})\partial_{12}^{2}G_{a+-}(s',s)\partial_{12}^{2}G_{b+-}(s',s)\right.\nonumber\\
&&+\left.u_{ba}(Q_{-})v_{ab}(Q_{-})\omega_{a}\omega_{b}\partial_{2}G_{a--}(s',s)\partial_{2}G_{b--}(s',s)+u_{ba}(Q_{-})u_{ab}(Q_{-})\partial_{12}^{2}G_{a--}(s',s)\partial_{12}^{2}G_{b--}(s',s)\right].\nonumber
\end{eqnarray}
}

At this point, it is useful to point out some further properties of the cavity mode  Green functions. Considering from Eq.(\ref{munukernels}) that $\mu'_{n}=-\omega_{n}\nu_{n}$, $\nu'_{n}=-\omega_{n}\mu_{n}$, $\mu''_{n}=-\omega_{n}^{2}\mu_{n}$ and $\nu''_{n}=-\omega_{n}^{2}\nu$, we can prove that:
\begin{equation}
\partial_{12}^{2}G_{n\pm\pm}(s',s)=\omega_{n}^{2}G_{n\pm\pm}(s',s)\mp\hbar\left[2\omega_{n}\nu_{n}(s'-s)\delta(s'-s)-\mu_{n}(s'-s)\delta'(s'-s)\right],
\end{equation}
\begin{equation}
\partial_{1}G_{n\pm\pm}(s',s)=\mp i\omega_{n}{\rm sgn}(s'-s)G_{n\mp\mp}(s',s)\mp\hbar\mu_{n}(s'-s)\delta(s'-s)=-\partial_{2}G_{n\pm\pm}(s',s).
\end{equation}
\begin{equation}
\partial_{12}^{2}G_{n\pm\mp}(s',s)=\omega_{n}^{2}G_{n\pm\mp}(s',s)~~,~~\partial_{1}G_{n\pm\mp}(s',s)=\pm i\omega_{n}G_{n\mp\pm}(s',s)=-\partial_{2}G_{n\pm\mp}(s',s).
\end{equation}

Then, we can split the terms according to the terms involving $G$'s and the remaining terms coming from the other terms containing delta functions and its derivative:
\begin{eqnarray}
&&S_{\rm IF}^{(2)}\left[Q_{+},Q_{-}\right]=-\frac{i}{\hbar}\int_{0}^{t}ds\int_{0}^{t}ds'\sum_{a,b}\omega_{a}^{2}\omega_{b}^{2}\\
&&\left[v_{ba}(Q_{+})G_{a++}(s',s)G_{b++}(s',s)v_{ab}(Q_{+})-v_{ba}(Q_{+})G_{a--}(s',s)G_{b--}(s',s)u_{ab}(Q_{+})\right.\nonumber\\
&&-\left.v_{ba}(Q_{+})G_{a-+}(s',s)G_{b-+}(s',s)v_{ab}(Q_{-})+v_{ba}(Q_{+})G_{a+-}(s',s)G_{b+-}(s',s)u_{ab}(Q_{-})\right.\nonumber\\
&&-\left.u_{ba}(Q_{+})G_{a--}(s',s)G_{b--}(s',s)v_{ab}(Q_{+})+u_{ba}(Q_{+})G_{a++}(s',s)G_{b++}(s',s)u_{ab}(Q_{+})\right.\nonumber\\
&&+\left.u_{ba}(Q_{+})G_{a+-}(s',s)G_{b+-}(s',s)v_{ab}(Q_{-})-u_{ba}(Q_{+})G_{a-+}(s',s)G_{b-+}(s',s)u_{ab}(Q_{-})\right.\nonumber\\
&&-\left.v_{ba}(Q_{-})G_{a+-}(s',s)G_{b+-}(s',s)v_{ab}(Q_{+})+v_{ba}(Q_{-})G_{a-+}(s',s)G_{b-+}(s',s)u_{ab}(Q_{+})\right.\nonumber\\
&&+\left.v_{ba}(Q_{-})G_{a--}(s',s)G_{b--}(s',s)v_{ab}(Q_{-})-v_{ba}(Q_{-})G_{a++}(s',s)G_{b++}(s',s)u_{ab}(Q_{-})\right.\nonumber\\
&&+\left.u_{ba}(Q_{-})G_{a-+}(s',s)G_{b-+}(s',s)v_{ab}(Q_{+})-u_{ba}(Q_{-})G_{a+-}(s',s)G_{b+-}(s',s)u_{ab}(Q_{+})\right.\nonumber\\
&&-\left.u_{ba}(Q_{-})G_{a++}(s',s)G_{b++}(s',s)v_{ab}(Q_{-})+u_{ba}(Q_{-})G_{a--}(s',s)G_{b--}(s',s)u_{ab}(Q_{-})\right]\nonumber\\
&&+{\text{(Remaining terms)}},\nonumber
\end{eqnarray}
making use of $v_{nm}=v_{mn}$ and $u_{nm}=u_{mn}$, so:
{
\fontsize{10pt}{12pt}\selectfont
\begin{eqnarray}
&&S_{\rm IF}^{(2)}\left[Q_{+},Q_{-}\right]=-\frac{i}{\hbar}\int_{0}^{t}ds\int_{0}^{t}ds'\sum_{a,b}\omega_{a}^{2}\omega_{b}^{2}\\
&&\left[\left(v_{ab}(Q_{+})v_{ab}(Q_{+})+u_{ab}(Q_{+})u_{ab}(Q_{+})-u_{ab}(Q_{-})v_{ab}(Q_{-})-v_{ab}(Q_{-})u_{ab}(Q_{-})\right)G_{a++}(s',s)G_{b++}(s',s)\right.\nonumber\\
&&-\left.\left(v_{ab}(Q_{+})v_{ab}(Q_{-})+u_{ab}(Q_{+})u_{ab}(Q_{-})-v_{ab}(Q_{-})u_{ab}(Q_{+})-u_{ab}(Q_{-})v_{ab}(Q_{+})\right)G_{a-+}(s',s)G_{b-+}(s',s)\right.\nonumber\\
&&-\left.\left(v_{ab}(Q_{-})v_{ab}(Q_{+})+u_{ab}(Q_{-})u_{ab}(Q_{+})-v_{ab}(Q_{+})u_{ab}(Q_{-})-u_{ab}(Q_{+})v_{ab}(Q_{-})\right)G_{a+-}(s',s)G_{b+-}(s',s)\right.\nonumber\\
&&+\left.\left(v_{ab}(Q_{-})v_{ab}(Q_{-})+u_{ab}(Q_{-})u_{ab}(Q_{-})-u_{ab}(Q_{+})v_{ab}(Q_{+})-v_{ab}(Q_{+})u_{ab}(Q_{+})\right)G_{a--}(s',s)G_{b--}(s',s)\right]\nonumber\\
&&+{\text{(Remaining terms)}}.\nonumber
\end{eqnarray}
}

At this point, it is crucial to recall the properties of the propagators. Considering that $G_{n++}$ corresponds to a Feynman propagator for the $n-$th oscillator, in such a way that $G_{n++}(s',s)=\theta(s'-s)G_{n-+}(s',s)+\theta(s-s')G_{n+-}(s',s)$, which stands as a time-ordered propagator (see Ref.\cite{CalHu08}). From this property, it is simple to observe that the product of two of these propagators is also a time-ordered function $G_{a++}(s',s)G_{b++}(s',s)=\theta(s'-s)G_{a-+}(s',s)G_{b-+}(s',s)+\theta(s-s')G_{a+-}(s',s)G_{b+-}(s',s)$.

As an example, we can take the terms including products of $v_{ab}$. Defining the convenient sum and difference variables $\Delta v_{ab}\equiv v_{ab}(Q_{+})-v_{ab}(Q_{-})$ and $\Sigma v_{ab}=[v_{ab}(Q_{+})+v_{ab}(Q_{-})]/2$, which when inverted gives $v_{ab}(Q_{+})=\Sigma v_{ab}+\Delta v_{ab}/2$ and $v_{ab}(Q_{-})=\Sigma v_{ab}-\Delta v_{ab}/2$, we can write:
\begin{eqnarray}
&&v_{ab}(Q_{+})v_{ab}(Q_{+})G_{a++}(s',s)G_{b++}(s',s)-v_{ab}(Q_{+})v_{ab}(Q_{-})G_{a-+}(s',s)G_{b-+}(s',s)\nonumber\\
&&-v_{ab}(Q_{-})v_{ab}(Q_{+})G_{a+-}(s',s)G_{b+-}(s',s)+v_{ab}(Q_{-})v_{ab}(Q_{-})G_{a--}(s',s)G_{b--}(s',s)=\nonumber\\
&&=\frac{1}{2}\left[G_{a-+}(s',s)G_{b-+}(s',s)+G_{a+-}(s',s)G_{b+-}(s',s)\right]\Delta v_{ab}(s)\Delta v_{ab}(s')\nonumber\\
&&+\theta(s'-s)\left[G_{a-+}(s',s)G_{b-+}(s',s)-G_{a+-}(s',s)G_{b+-}(s',s)\right]\Sigma v_{ab}(s)\Delta v_{ab}(s')\nonumber\\
&&+\theta(s-s')\left[G_{a+-}(s',s)G_{b+-}(s',s)-G_{a-+}(s',s)G_{b-+}(s',s)\right]\Delta v_{ab}(s)\Sigma v_{ab}(s'),
\end{eqnarray}
where the $\Sigma v_{ab}(s)\Sigma v_{ab}(s')$ vanishes since it can be proven that $G_{a++}G_{b++}-G_{a-+}G_{b-+}-G_{a+-}G_{b+-}+G_{a--}G_{b--}=0$. Moreover, considering that $G_{n+-}(t,t')=G_{n-+}(t',t)$ and by switching $s'\leftrightarrow s$ in some of the terms, we obtain:
\begin{eqnarray}
&&v_{ab}(Q_{+})v_{ab}(Q_{+})G_{a++}(s',s)G_{b++}(s',s)-v_{ab}(Q_{+})v_{ab}(Q_{-})G_{a-+}(s',s)G_{b-+}(s',s)\nonumber\\
&&-v_{ab}(Q_{-})v_{ab}(Q_{+})G_{a+-}(s',s)G_{b+-}(s',s)+v_{ab}(Q_{-})v_{ab}(Q_{-})G_{a--}(s',s)G_{b--}(s',s)=\nonumber\\
&&=\Delta v_{ab}(s)\frac{1}{2}\left[G_{a-+}(s,s')G_{b-+}(s,s')+G_{a+-}(s,s')G_{b+-}(s,s')\right]\Delta v_{ab}(s')\nonumber\\
&&+\Delta v_{ab}(s)2\theta(s-s')\left[G_{a-+}(s,s')G_{b-+}(s,s')-G_{a+-}(s,s')G_{b+-}(s,s')\right]\Sigma v_{ab}(s')\nonumber\\
&&=i\hbar\left[-\Delta v_{ab}(s)2\frac{\mathcal{D}_{ab}(s,s')}{\omega_{a}^{2}\omega_{b}^{2}}\Sigma v_{ab}(s')+\Delta v_{ab}(s)\frac{i}{2}\frac{\mathcal{N}_{ab}(s,s')}{\omega_{a}^{2}\omega_{b}^{2}}\Delta v_{ab}(s')\right],
\end{eqnarray}
where the dissipation and noise kernels read:
\begin{eqnarray}
\frac{\hbar\mathcal{D}_{ab}(s,s')}{\omega_{a}^{2}\omega_{b}^{2}}&=&i\theta(s-s')\left[G_{a-+}(s,s')G_{b-+}(s,s')-G_{a+-}(s,s')G_{b+-}(s,s')\right]\nonumber\\
&=&-\theta(s-s')2\hbar^{2}\left[\mu_{a}(s-s')\nu_{b}(s-s')+\nu_{a}(s-s')\mu_{b}(s-s')\right]\nonumber\\
&=&\theta(s-s')2q_{a,{\rm zp}}^{2}q_{b,{\rm zp}}^{2}\sin\left[\left(\omega_{a}+\omega_{b}\right)(s-s')\right],
\end{eqnarray}
\begin{eqnarray}
\frac{\hbar\mathcal{N}_{ab}(s,s')}{\omega_{a}^{2}\omega_{b}^{2}}&=&-G_{a+-}(s,s')G_{b+-}(s,s')-G_{a-+}(s,s')G_{b-+}(s,s')\nonumber\\
&=&2\hbar^{2}\left[\nu_{a}(s-s')\mu_{b}(s-s')-\mu_{a}(s-s')\mu_{b}(s-s')\right]\nonumber\\
&=&2q_{a,{\rm zp}}^{2}q_{b,{\rm zp}}^{2}\cos\left[\left(\omega_{a}+\omega_{b}\right)(s-s')\right],
\end{eqnarray}
where both kernels are real, the dissipation kernel is a causal function while the noise kernel is symmetric in the exchange $s\leftrightarrow s'$. The same combination is obtained for the terms including the products of $u_{ab}$.

The cross terms between $v_{ab}$ and $u_{ab}$  can be worked out similarly. While the $\Sigma\Sigma$ cross  terms always vanish, the $\Delta\Delta$ cross terms end up giving $-\mathcal{N}_{ab}$ and the $\Delta\Sigma$ and $\Sigma\Delta$ cross terms can be grouped together giving the dissipation kernel $-\mathcal{D}_{ab}$. All in all, the terms including each kernel can be grouped, leading to:
\begin{eqnarray}
S_{\rm IF}^{(2)}\left[Q_{+},Q_{-}\right]&=&\int_{0}^{t}ds\int_{0}^{t}ds'\sum_{a,b}\left(-\left[\Delta v_{ab}(s)-\Delta u_{ab}(s)\right]2\mathcal{D}_{ab}(s,s')\left[\Sigma v_{ab}(s')-\Sigma u_{ab}(s')\right]\right.\\
&&+\left.\left[\Delta v_{ab}(s)-\Delta u_{ab}(s)\right]\frac{i}{2}\mathcal{N}_{ab}(s,s')\left[\Delta v_{ab}(s')-\Delta u_{ab}(s')\right]\right)+{\text{(Remaining terms)}}.\nonumber
\end{eqnarray}

The remaining terms are:
\begin{eqnarray}
&&{\text{(Remaining terms)}}=-\frac{i}{\hbar}\int_{0}^{t}ds\int_{0}^{t}ds'\sum_{a,b}\label{RemainingTermsInitial}\\
&&\left(v_{ab}(Q_{+})\omega_{b}\omega_{a}u_{ab}(Q_{+})\left[\partial_{1}G_{a++}(s',s)\partial_{1}G_{b++}(s',s)+\omega_{a}\omega_{b}G_{a--}(s',s)G_{b--}(s',s)\right]\right.\nonumber\\
&&+\left.v_{ab}(Q_{-})\omega_{b}\omega_{a}u_{ab}(Q_{-})\left[\partial_{1}G_{a--}(s',s)\partial_{1}G_{b--}(s',s)+\omega_{a}\omega_{b}G_{a++}(s',s)G_{b++}(s',s)\right]\right.\nonumber\\
&&+\left.u_{ab}(Q_{+})v_{ab}(Q_{+})\omega_{a}\omega_{b}\left[\partial_{2}G_{a++}(s',s)\partial_{2}G_{b++}(s',s)+\omega_{a}\omega_{b}G_{a--}(s',s)G_{b--}(s',s)\right]\right.\nonumber\\
&&+\left.u_{ab}(Q_{-})v_{ab}(Q_{-})\omega_{a}\omega_{b}\left[\partial_{2}G_{a--}(s',s)\partial_{2}G_{b--}(s',s)+\omega_{a}\omega_{b}G_{a++}(s',s)G_{b++}(s',s)\right]\right.\nonumber\\
&&+\left.u_{ab}(Q_{+})u_{ab}(Q_{+})\left[\partial_{12}^{2}G_{a++}(s',s)\partial_{12}^{2}G_{b++}(s',s)-\omega_{a}^{2}\omega_{b}^{2}G_{a++}(s',s)G_{b++}(s',s)\right]\right.\nonumber\\
&&+\left.u_{ab}(Q_{-})u_{ab}(Q_{-})\left[\partial_{12}^{2}G_{a--}(s',s)\partial_{12}^{2}G_{b--}(s',s)-\omega_{a}^{2}\omega_{b}^{2}G_{a--}(s',s)G_{b--}(s',s)\right]\right).\nonumber
\end{eqnarray}

Take the first term:
\begin{eqnarray}
\partial_{1}G_{a++}(s',s)\partial_{1}G_{b++}(s',s)&+&\omega_{a}\omega_{b}G_{a--}(s',s)G_{b--}(s',s)=\\
&=&i\omega_{a}{\rm sgn}(s'-s)G_{a--}(s',s)\hbar\mu_{b}(s'-s)\delta(s'-s)\nonumber\\
&&+\hbar\mu_{a}(s'-s)\delta(s'-s)i\omega_{b}{\rm sgn}(s'-s)G_{b--}(s',s)\nonumber\\
&&+\hbar^{2}\mu_{a}(s'-s)\delta(s'-s)\mu_{b}(s'-s)\delta(s'-s).\nonumber
\end{eqnarray}

When it is placed into Eq.(\ref{RemainingTermsInitial}), it collapses one of the integrals with $\delta(s'-s)$. We immediately obtain factors $\mu_{n}(0)=0$ which anul the contribution. Similarly, this happens for the terms involving $\partial_{1,2}G_{a,b,\pm\pm}$. Then, a first simplification results:
\begin{eqnarray}
&&{\text{(Remaining terms)}}=-\frac{i}{\hbar}\int_{0}^{t}ds\int_{0}^{t}ds'\sum_{a,b}\label{RemainingTermsSimplified}\\
&&\times\left(u_{ab}(Q_{+})u_{ab}(Q_{+})\left[\partial_{12}^{2}G_{a++}(s',s)\partial_{12}^{2}G_{b++}(s',s)-\omega_{a}^{2}\omega_{b}^{2}G_{a++}(s',s)G_{b++}(s',s)\right]\right.\nonumber\\
&&+\left.u_{ab}(Q_{-})u_{ab}(Q_{-})\left[\partial_{12}^{2}G_{a--}(s',s)\partial_{12}^{2}G_{b--}(s',s)-\omega_{a}^{2}\omega_{b}^{2}G_{a--}(s',s)G_{b--}(s',s)\right]\right).\nonumber
\end{eqnarray}

These terms are given by:
\begin{eqnarray}
&&\partial_{12}^{2}G_{a\pm\pm}(s',s)\partial_{12}^{2}G_{b\pm\pm}(s',s)-\omega_{a}^{2}\omega_{b}^{2}G_{a\pm\pm}(s',s)G_{b\pm\pm}(s',s)=\\
&&=\mp\omega_{a}^{2}G_{a\pm\pm}(s',s)\hbar\left[ 2\omega_{b}\nu_{b}(s'-s)\delta(s'-s)-\mu_{b}(s'-s)\delta'(s'-s)\right]\nonumber\\
&&\mp\hbar\left[2\omega_{a}\nu_{a}(s'-s)\delta(s'-s)-\mu_{a}(s'-s)\delta'(s'-s)\right]\omega_{b}^{2}G_{b\pm\pm}(s',s)\nonumber\\
&&+\hbar^{2}\left[2\omega_{b}\nu_{b}(s'-s)\delta(s'-s)-\mu_{b}(s'-s)\delta'(s'-s)\right]\left[2\omega_{a}\nu_{a}(s'-s)\delta(s'-s)-\mu_{a}(s'-s)\delta'(s'-s)\right].\nonumber
\end{eqnarray}

Observing which terms are the ones contributing and simplifying them by the action of the delta functions, we get:
\begin{eqnarray}
&&\partial_{12}^{2}G_{a\pm\pm}(s',s)\partial_{12}^{2}G_{b\pm\pm}(s',s)-\omega_{a}^{2}\omega_{b}^{2}G_{a\pm\pm}(s',s)G_{b\pm\pm}(s',s)=\nonumber\\
&&=\mp\omega_{a}^{2}G_{a\pm\pm}(s',s)\hbar 2\omega_{b}\nu_{b}(s'-s)\delta(s'-s)\pm\omega_{a}^{2}G_{a\pm\pm}(s',s)\hbar\mu_{b}(s'-s)\delta'(s'-s)\\
&&\mp\hbar 2\omega_{a}\nu_{a}(s'-s)\delta(s'-s)\omega_{b}^{2}G_{b\pm\pm}(s',s)\pm\hbar\mu_{a}(s'-s)\delta'(s'-s)\omega_{b}^{2}G_{b\pm\pm}(s',s)\nonumber\\
&&+\hbar^{2}2\omega_{b}\nu_{b}(s'-s)\delta(s'-s)2\omega_{a}\nu_{a}(s'-s)\delta(s'-s)+\hbar^{2}\mu_{b}(s'-s)\delta'(s'-s)\mu_{a}(s'-s)\delta'(s'-s).\nonumber\\
&&=\mp\omega_{a}^{2}G_{a\pm\pm}(0)\hbar 2\omega_{b}\nu_{b}(0)\delta(s'-s)\pm\omega_{a}^{2}G_{a\pm\pm}(s',s)\hbar\mu_{b}(s'-s)\delta'(s'-s)\\
&&\mp\hbar 2\omega_{a}\nu_{a}(0)\delta(s'-s)\omega_{b}^{2}G_{b\pm\pm}(0)\pm\hbar\mu_{a}(s'-s)\delta'(s'-s)\omega_{b}^{2}G_{b\pm\pm}(s',s)\nonumber\\
&&+\hbar^{2}2\omega_{b}\nu_{b}(0)\delta(0)2\omega_{a}\nu_{a}(0)\delta(s'-s)+\hbar^{2}\mu_{b}(s'-s)\delta'(s'-s)\mu_{a}(s'-s)\delta'(s'-s).\nonumber\\
&&=2i\left[\mp(\omega_{a}+\omega_{b})-2i\delta(0)\right]\omega_{a}\omega_{b}q_{a,{\rm zp}}^{2}q_{b,{\rm zp}}^{2}\delta(s'-s)\\
&&\pm\hbar\left[\mu_{a}(s'-s)\omega_{b}^{2}G_{b\pm\pm}(s',s)+\mu_{b}(s'-s)\omega_{a}^{2}G_{a\pm\pm}(s',s)\right]\delta'(s'-s)\nonumber\\
&&+\hbar^{2}\mu_{b}(s'-s)\delta'(s'-s)\mu_{a}(s'-s)\delta'(s'-s).\nonumber
\end{eqnarray}

Inserting into Eq.(\ref{RemainingTermsSimplified}), and grouping the terms including delta functions we have:
\begin{eqnarray}
&&-\frac{i}{\hbar}\int_{0}^{t}ds\int_{0}^{t}ds'\sum_{a,b}u_{ab}(Q_{\pm})u_{ab}(Q_{\pm})\left[\partial_{12}^{2}G_{a\pm\pm}(s',s)\partial_{12}^{2}G_{b\pm\pm}(s',s)-\omega_{a}^{2}\omega_{b}^{2}G_{a\pm\pm}(s',s)G_{b\pm\pm}(s',s)\right]=\nonumber\\
&&=-\frac{i}{\hbar}\int_{0}^{t}ds\int_{0}^{t}ds'\sum_{a,b}u_{ab}(Q_{\pm})u_{ab}(Q_{\pm})\left(2i\left[\mp(\omega_{a}+\omega_{b})-2i\delta(0)\right]\omega_{a}\omega_{b}q_{a,{\rm zp}}^{2}q_{b,{\rm zp}}^{2}\delta(s'-s)\right.\\
&&\pm\left.\hbar\left[\mu_{a}(s'-s)\omega_{b}^{2}G_{b\pm\pm}(s',s)+\mu_{b}(s'-s)\omega_{a}^{2}G_{a\pm\pm}(s',s)\right]\delta'(s'-s)\right.\nonumber\\
&&+\left.\hbar^{2}\mu_{b}(s'-s)\delta'(s'-s)\mu_{a}(s'-s)\delta'(s'-s)\right)\nonumber\\
&&=\frac{1}{\hbar}\int_{0}^{t}ds\sum_{a,b}u_{ab}(Q_{\pm})u_{ab}(Q_{\pm})\left[\mp(\omega_{a}+\omega_{b})-4i\delta(0)\right]\omega_{a}\omega_{b}q_{a,{\rm zp}}^{2}q_{b,{\rm zp}}^{2}.\nonumber
\end{eqnarray}

Considering the last expression, we have that:
\begin{eqnarray}
&&S_{\rm IF}^{(2)}\left[Q_{+},Q_{-}\right]=\int_{0}^{t}ds\left(-\sum_{a,b}\delta V_{ab+}^{(2)}(Q_{+})\right)-\int_{0}^{t}ds\left(-\sum_{a,b}\delta V_{ab-}^{(2)}(Q_{-})\right)\\
&&+\int_{0}^{t}ds\int_{0}^{t}ds'\sum_{a,b}\Big(-\left[\Delta v_{ab}(s)-\Delta u_{ab}(s)\right]2\mathcal{D}_{ab}(s,s')\left[\Sigma v_{ab}(s')-\Sigma u_{ab}(s')\right]\nonumber\\
&&+\left[\Delta v_{ab}(s)-\Delta u_{ab}(s)\right]\frac{i}{2}\mathcal{N}_{ab}(s,s')\left[\Delta v_{ab}(s')-\Delta u_{ab}(s')\right]\Big),\nonumber
\end{eqnarray}
where the correction to the potentials read:
\begin{equation}
\delta V_{ab\pm}^{(2)}(Q_{\pm})=\frac{1}{\hbar}\omega_{a}\omega_{b}q_{a,{\rm zp}}^{2}q_{b,{\rm zp}}^{2}\left[\omega_{a}+\omega_{b}\pm 4i\delta(0)\right]u_{ab}(Q_{\pm})u_{ab}(Q_{\pm}).
\end{equation}

\end{document}